\documentclass[twocolumn]{aastex63}

\usepackage{amssymb}
\usepackage{graphicx}
\usepackage{verbatim}
\usepackage{datetime}
\usepackage{nicefrac}

\newcommand{\HI}{{\ion{H}{1}}}

\newcommand{\matHI}{\rm H{\hskip 0.02cm\scriptscriptstyle I}}

\newcommand{\n}{\nodata}
\newcommand{\be}{\begin{itemize}}
\newcommand{\ee}{\end{itemize}}
\newcommand{\muasyr}{\hbox{$\; \mu{\rm as \ y}^{-1}\;$}}

\newcommand{\chandra}{{\it{}Chandra}\ }
\newlength\mystoreparindent

\def\fermi{\textit{Fermi }}
\def\gr{$\gamma$-ray }

\received{30 April 2020}
\revised{15 June 2020}
\accepted{29 June 2020}

\submitjournal{Astrophysical Journal}
 
\shorttitle{TXS~0128+554}
\shortauthors{M. L. Lister et al.}

\begin{document}

\title{TXS~0128+554: A Young Gamma-Ray Emitting AGN With Episodic Jet Activity}

\correspondingauthor{M. L. Lister}
\email{mlister@purdue.edu}

\author{M. L. Lister}
\affiliation{Department of Physics and Astronomy, Purdue University, 525 Northwestern Avenue, West Lafayette, IN 47907, USA}

\author{D. C. Homan}
\affiliation{Department of Physics, Denison University, Granville, OH 43023, USA}

\author{Y. Y. Kovalev}
\affiliation{Astro Space Center of Lebedev Physical Institute, Profsoyuznaya 84/32, 117997 Moscow, Russia}
\affiliation{Moscow Institute of Physics and Technology, Institutsky per. 9, Dolgoprudny 141700, Russia}
\affiliation{Max-Planck-Institut f\"ur Radioastronomie, Auf dem H\"ugel 69, 53121 Bonn, Germany}

\author{S. Mandal}
\affiliation{Department of Physics and Astronomy, Purdue University, 525 Northwestern Avenue, West Lafayette, IN 47907, USA}

\author{A. B. Pushkarev}
\affiliation{Crimean Astrophysical Observatory, Nauchny 298688, Crimea, Russia}
\affiliation{Astro Space Center of Lebedev Physical Institute, Profsoyuznaya 84/32, 117997 Moscow, Russia}

\author{A. Siemiginowska}
\affiliation{Center for Astrophysics $|$ Harvard and Smithsonian, 60 Garden St., Cambridge, MA 02138, USA}

\begin{abstract}

We have carried out a {\it Chandra} X-ray and multi-frequency radio VLBA study of the AGN TXS~0128+554, which is associated
with the \fermi \gr source 4FGL J0131.2+5547. The AGN is unresolved in a target 19.3 ks {\it Chandra} image, and its spectrum is well fit by a simple absorbed power law model, with no distinguishable spectral features. Its relatively soft X-ray spectrum compared to other CSOs may be indicative of  a thermal emission component, for which we were
able to obtain an upper temperature limit of kT = 0.08 keV. The compact radio morphology and measured advance speed of $0.32\, c \pm 0.07\, c$ indicate a kinematic age of only 82 y $\pm 17$ y, placing TXS~0128+554 among the youngest members of the compact symmetric object (CSO) class. The lack of compact, inverted spectrum hotspots and an emission gap between the bright inner jet and outer radio lobe structure indicate that the jets have undergone episodic activity, and were re-launched a decade ago. The predicted \gr emission from the lobes, based on an inverse Compton-emitting cocoon model, is three orders of magnitude below the observed \fermi LAT flux. A comparison to other \textit{Fermi}-detected and non-\textit{Fermi} detected CSOs with redshift $z < 0.1$ indicates that the \gr emission likely originates in the inner jet/core region, and that nearby, recently launched AGN jets are primary candidates for detection by the \fermi LAT instrument. 
\\
\end{abstract}

%
 
\section{Introduction}

The \fermi LAT (4LAC) catalog of high-confidence AGN associations published by \cite{4LAC} contains 2863 \gr source detections above 0.1 GeV, 98\% of which are classified as blazars or blazar candidates.  These AGN have powerful relativistic jetted outflows aligned  close to our line of sight, and dominate the \gr sky due to strong Doppler boosting of their high-energy emission \citep[e.g.,][]{MF1,MF2,2010AA...512A..24S}.

At present there are only a few known detections of weakly-boosted \gr
emission from non-blazar (i.e., ``misaligned'') AGN. Of the dozen \gr
loud radio galaxies listed by \cite{2016arXiv161102986R}, all but 3
show either one-sided pc-scale jets and/or superluminal jet motion,
indicative of small viewing angles.  The exceptions are Centaurus A
(distance d = 4 Mpc) and Fornax A (d = 18 Mpc), where resolved \gr
emission has been detected from their kpc-scale lobes, and 3C~84 (d =
70 Mpc, in the Perseus cluster).  The latter is unusual in terms of
having two-sided radio jet structure on both pc- and kpc-scales, a
strong compact radio core, and rapidly variable \gr emission
\citep{2009ApJ...699...31A}.

The \fermi catalog statistics suggest that extra-galactic \gr emission can only be
detected in either highly boosted jets, or in very nearby radio
galaxies. However, there have been theoretical predictions of
strong \gr emission from the pc-scale lobes of powerful young radio
sources. These lobes are powered by recently activated jets that are plowing into dense gas in their host galaxy
\citep{2008ApJ...680..911S,2009MNRAS.395L..43K, 2007MNRAS.376.1630K,2011MNRAS.412L..20K}.  Very few such candidates
have been identified to date. \cite{2014ApJ...780..165M} initially
found no \gr counterparts at the locations of twelve young radio sources,
but subsequently claimed a LAT detection near the position of PKS
1718--649 \citep{2016ApJ...821L..31M}. The latter is believed to have a
kinematic age of less than 100 years \citep{2009AN....330..193G}.  Also
\cite{2011ApJ...738..148M} and \cite{2014AA...562A...4M} have claimed
that the \gr AGN IERS B0954+556 and PMN~J1603$-$4904, respectively, may
either be young radio sources, or unusual examples of beamed blazars. 
Finally, \cite{2020AA...635A.185P} have reported another young radio galaxy, NGC 3894, as a high-confidence LAT detection. 

Here we present the results of a multi-frequency VLBA study of the
nearby ($z = 0.0365$; \citealt{2012ApJS..199...26H}) AGN TXS~0128+554, which is associated with the \fermi \gr catalog source 4FGL J0131.2+5547
/ 3FHL J0131.1+5546. The radio source has a compact morphology that indicates it is a new example of a young, non-blazar AGN that is emitting at \gr
energies. It is among the lowest luminosity \gr AGN detected by \fermi to date. We also present targeted 19~ks Chandra X-ray observations, and analyze the overall measurements in terms of the \cite{2008ApJ...680..911S} and \cite{2009MNRAS.395L..43K} cocoon models. We compare the physical characteristics of TXS~0128+554 to those of other known young AGN, and conclude that the \fermi LAT \gr detection of these radio sources depends strongly on both their cosmological distance and the current activity state of their inner jet region. 

Throughout this paper we adopt the convention $S_\nu \propto
\nu^{\alpha}$ for spectral index $\alpha$, and use the cosmological
parameters $\Omega_m = 0.27$, $\Omega_\Lambda = 0.73$ and $H_o = 71 \;
\mathrm{km\; s^{-1} \; Mpc^{-1}}$ \citep{Komatsu09}. Under these
assumptions the redshift of TXS~0128+554 corresponds to a luminosity
distance of 158 Mpc and a linear scaling of 0.72 pc per
milliarcsecond. With regards to the \fermi \gr observatory, we note that true \fermi AGN detections are rare, due to the poor angular localization of sources on the sky by the LAT instrument. They typically require additional information, such as co-temporaneous multi-wavelength variability. At the same time, the \fermi associations made on the basis of VLBI catalogs are shown to deliver a very low probability of chance coincidence \citep{2009ApJ...707L..56K}. For the purposes of this paper, we will refer to AGN listed as high-confidence associations in the \fermi LAT catalogs as \fermi detections. 

\section{\label{background} Background and Identification}

MOJAVE (Monitoring of Jets in AGN with VLBA Experiments) is a long
term project to study the kinematic and polarization evolution of
pc-scale AGN jets with regular VLBA imaging at 15 GHz
\citep{MOJAVE_V}.  It originally focused on a complete sample of the
135 strongest, most compact radio-loud AGN in the northern sky, and
has been expanded during the \fermi era to encompass over 400
AGN.  In September 2016, the survey added all AGN in the 3FGL catalog
with mean \gr photon index harder than 2.1 and an associated radio
source with at least 100 mJy of flux density at 15 GHz. The majority
of these have a spectral energy distribution with a synchrotron peak
above $10^{13}$ Hz, placing them in the intermediate- or
high-spectral peaked classes. They are thus prime candidates for
detection at TeV \gr energies with current (e.g., VERITAS, HESS, HAWC)
and future high-energy observatories such as CTA.

The first VLBA images of these sources were obtained in late 2016, and
show typical one-sided (core + jet) structures, similar to other AGN
in the MOJAVE survey \citep{MOJAVE_XV}. However, the radio galaxy TXS~0128+554 stood out
as having a two-sided (lobe-core-lobe) morphology similar to known
compact symmetric objects (CSOs).  This class was first identified in
the 1990s and generally consists of young radio sources that
enter into flux-density-limited surveys due to their powerful
pc-scale lobe emission \citep{1998PASP..110..493O}.  Because they are
not selected on the basis of Doppler boosted jet emission, they are
not subject to orientation bias, and therefore are more likely have jets lying close
to the plane of the sky.  Their lobe emission is usually confined to
pc-scales, and typically dominates over that of the core, resulting in
either a convex spectrum, or one that remains steep down to MHz
frequencies.  The radio emission from GPS/CSO jets is often strongly depolarized due to dense ionized gas in their external environment \citep{2003ApJ...586...33A,2013MNRAS.433..147D, 2016MNRAS.459..820T}.  

\begin{deluxetable*}{cccccccc} 
\tablewidth{0pt}  
\tablecaption{\label{t:imagestats} VLBA Image Properties}
\tablehead{
\colhead{$\nu$} & \colhead{$I_\mathrm{tot}$}   &\colhead{$I_\mathrm{peak}$} &  \colhead{Maj.} &   \colhead{Min.} &   \colhead{PA}\ &  \colhead{$I_\mathrm{rms}$} & \colhead{$I_\mathrm{base}$} \\[-2ex] 
\colhead{(GHz)} & \colhead{(mJy)} & \colhead{(mJy$\ \mathrm{bm}^{-1}$)} & \colhead{(mas)} & \colhead{(mas)} & \colhead{($^\circ$)} & \colhead{(mJy$\ \mathrm{bm}^{-1}$)} & \colhead{(mJy$\ \mathrm{bm}^{-1}$)}
}
\decimalcolnumbers
\startdata 
2.3	&207	&79	&5.71	&4.42	&$-$10.2	&0.07	&0.30 \\
5	&162	&61	&2.21	&1.87	&0.3	&0.03	&0.10 \\
6.6	&151	&62	&1.63	&1.37	&$-$3.1	&0.04	&0.15 \\
8.4	&151	&64	&1.35	&1.12	&$-$9.0	&0.05	&0.16 \\
15.4	&114	&63	&0.70	&0.58	&$-$11.8	&0.04	&0.12 \\
22.2	&97	&84	&0.53	&0.45	&$-$8.8	&0.06	&0.20 \\
\enddata
\tablecomments{Columns are as follows: (1) observing frequency (GHz),
  (2) total cleaned flux density (mJy), (3) map peak (mJy per beam),
  (4) FWHM major axis of restoring beam (milliarcseconds), (5) FWHM
  minor axis of restoring beam (milliarcseconds), (6) position angle
  of major axis of restoring beam (degrees), (7) rms noise level of
  image (mJy per beam), (11) lowest I contour (mJy per beam).}
\end{deluxetable*}

The radio source TXS~0128+554 is positionally coincident with the nucleus of a $\mathrm{K_s} = 10.7$ mag elliptical galaxy in the 2MASS survey at $z = 0.036$ (158 Mpc; \citealt{2012ApJS..199...26H}).  Its WISE infrared colors place it within the blazar color-color strip as defined by \cite{2012ApJ...750..138M} but well outside the region occupied by $\gamma$-ray-loud blazars. It is associated with the {\it ROSAT} X-ray
source RX J0131.2+5545, and the LAT source 4FGL J0131.2+5547. The radio position is offset 116 arcsec from the LAT centroid, but is within the 95\% confidence ellipsoid dimensions of 143 arcsec by 122 arcsec. The \gr source was listed in the \fermi 3LAC (high-confidence AGN) catalog \citep{3LAC} but not in 4LAC \citep{4LAC}. In the 4FGL catalog it is fitted by a single power law spectrum of index = 2.10$\pm$ 0.09 and an energy flux of 5.6 $\times 10^{-13}$ erg cm$^{-2}$ s$^{-1}$ between 0.1 GeV and 100 GeV (detection significance = 11 sigma). It has a \gr variability index of 62, which is below the value of 72 at which sources in the 4FGL are considered variable at $>99$\,\% confidence \citep{4FGL}. Its properties above 10 GeV are also tabulated in the 3FHL catalog \citep{3FHL}, with a relatively steep power law index of $3.1 \pm 0.9$ and no significant spectral curvature.  The source was listed as an association with 3FGL J0131.3+5548 in the 3FGL catalog \citep{3FGL}, which included \fermi data from 2008--2012, but not in any of the previous EGRET or \fermi \gr catalogs. No observations have been reported to date at TeV energies.

\section{\label{data}OBSERVATIONAL DATA}

\subsection{Multi-frequency VLBA observations}

We observed TXS~0128+554 over a 13 h period on 2018
June 29 (obscode BL251) with the VLBA. The North Liberty antenna did
not participate due to a damaged elevation motor.  Two scans were made
of the bright radio blazar TXS~0059+581 for fringe calibration. During
each of the target and calibrator pointings, scans were made
successively at central observing frequencies of 2.3 GHz, 5.0 GHz, 6.6
GHz, 8.4 GHz, 15.4 GHz, 22.2 GHz. The data were recorded at 2048 Mbps
in dual circular polarization, with a bandwidth of 256 MHz at each
frequency. 

\begin{figure*}[p]
\centering
\includegraphics[width=0.78\linewidth]{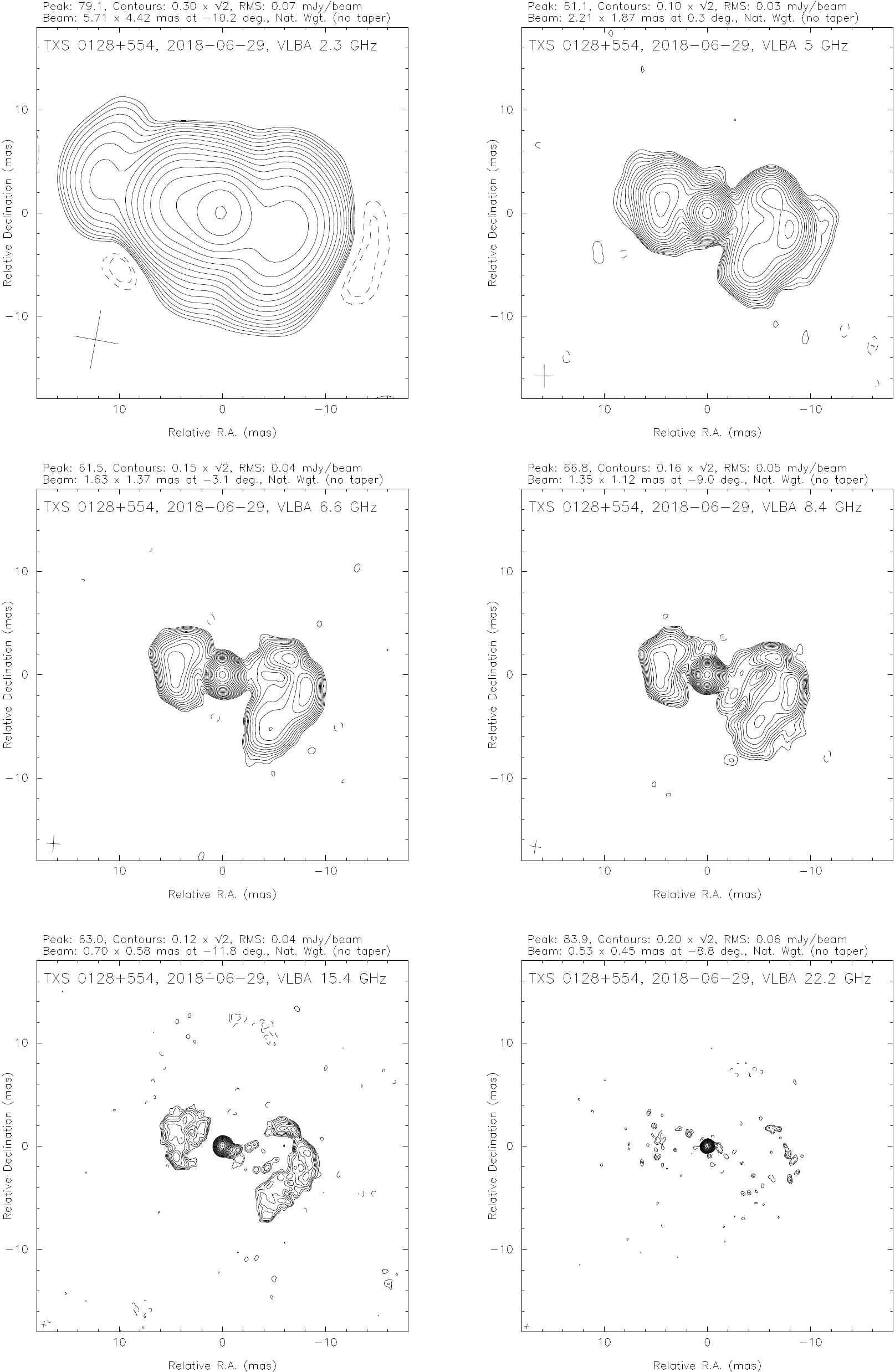}
\caption{\label{multifreqmap} Clockwise from the top left: total intensity VLBA
  contour maps  of TXS~0128$+$554 at 2.3 GHz, 5 GHz, 6.6 GHz, 8.4 GHz, 15.4
  GHz, and 22.2 GHz. The Gaussian FWHM restoring beam dimensions of each map are
  indicated by a cross in the lower left corner of each panel.  }
\end{figure*}

We processed the data in AIPS \citep{AIPS} following standard
procedures, and self-calibrated and imaged the data with the Difmap
package \citep{DIFMAP}. Approximately 40\% of the data at 2.3 GHz were
unusable due to radio frequency interference from satellite radio signals. In Figure~\ref{multifreqmap} we show the natural-weight total intensity contour maps at all six frequencies, and list the map properties in Table~\ref{t:imagestats}.

\begin{deluxetable}{clcrrcc} 
\tablewidth{0pt}  
\tablecaption{\label{t:gaussiantable}Fitted 15 GHz Jet Features}  
\tablehead{\colhead {} &   \colhead {} & 
 \colhead{I} & \colhead{r} &\colhead{P.A.} & \colhead{Maj.} & \colhead{log $T_b$} \\[-2ex]
\colhead {I.D.} &  \colhead {Epoch} & 
\colhead{(mJy)} & \colhead{(mas)} &\colhead{(\arcdeg)} & \colhead{(mas)} & \colhead{(K)}
} 
\decimalcolnumbers
\startdata 
C0& 2016 Sep 26  & 59.2  & 0.02 & 199.0 & 0.15 & 10.1 \\ 
C0& 2016 Nov 6  & 56.4  & \n & \n & 0.15 & 10.1 \\ 
C0& 2016 Dec 10  & 54.1  & \n & \n & 0.11 & 10.4 \\ 
C0& 2017 Jan 28  & 51.9  & \n & \n & 0.14 & 10.1 \\ 
C0& 2017 Jul 30  & 56.8  & \n & \n & 0.12 & 10.3 \\ 
C0& 2017 Nov 18  & 63.3  & \n & \n & 0.16 & 10.1 \\ 
C0& 2018 Jun 29  & 49.7  & 0.04 & 91.0 & \n & \n \\ 
C0& 2019 Aug 23  & 41.9  & 0.02 & 101.3 & 0.09 & 10.4 \\ 
C1& 2016 Sep 26  & 2.5  & 4.59 & 105.5 & 0.49 & 7.7 \\ 
C1& 2016 Nov 6  & 3.0  & 4.62 & 105.0 & 0.64 & 7.6 \\ 
C1& 2016 Dec 10  & 2.4  & 4.64 & 103.6 & 0.55 & 7.6 \\ 
C1& 2017 Jan 28  & 2.5  & 4.64 & 104.1 & 0.50 & 7.7 \\ 
C1& 2017 Jul 30  & 1.5  & 4.58 & 102.5 & \n & \n \\ 
C1& 2017 Nov 18  & 3.1  & 4.51 & 104.0 & 0.87 & 7.3 \\ 
C1& 2018 Jun 29  & 3.7  & 4.52 & 104.2 & 0.96 & 7.3 \\ 
C1& 2019 Aug 23  & 4.2  & 4.45 & 104.6 & 1.43 & 7.0 \\ 
\enddata 
\tablecomments{Columns are as follows:  (1) feature identification number (zero indicates core feature), (2) observation epoch, (3) flux density at 15 GHz in mJy, (4) position offset from the core feature (or map center for the core feature entries) in milliarcseconds, (5) position angle with respect to the core feature (or map center for the core feature entries) in degrees,  (6) FWHM major axis of fitted Gaussian in milliarcseconds, (7) log of brightness temperature in Kelvin.  (This table is available in its entirety in machine-readable form.)}
\end{deluxetable}

We fitted bright jet features in the visibility plane to the 15 GHz data and other MOJAVE 15 GHz epochs (\S~\ref{kinematics}) with circular Gaussians (Table~\ref{t:gaussiantable}). Based on
previous analysis \citep{MOJAVE_VI}, the typical uncertainties in the feature centroid positions are $\sim 20$\% of the FWHM naturally-weighted image restoring beam dimensions. For the bright and compact features, the positional errors are smaller by approximately a factor of two.  We estimate the formal errors on the feature sizes to be roughly twice the positional error, according to \cite{1999ASPC..180..301F}.  The flux density accuracies are approximately 5\% \citep{2002ApJ...568...99H}.

\subsection{X-ray Observations}

\begin{deluxetable}{crcc} 
    \tablewidth{0pt}  
    \tablecaption{\label{t:chandraobstable} Chandra Observations}
    \tablehead{ \colhead{Date}  & \colhead{ OBSID} & \colhead{Exposure} 
    &\colhead{Net Counts$^{a}$}\\[-2ex]
    & & \colhead{(ks)}}
\startdata
 2019 Mar 29 & 21408 & 5.76 & $400.4\pm20.6$\\
 2019 Mar 30 & 22160 & 6.68 & $528.2\pm23.1$\\
 2019 Mar 31 & 22161 & 3.84 & $301.5\pm17.0$\\
 2019 Apr 01 & 22162 & 3.06 & $257.9\pm16.1$\\
\enddata

\tablenotetext{a}{ Net counts given for a 
        circular source region with a radius $r=1.5$ arcec, with 
        1$\sigma$ errors.}
\end{deluxetable} 

The \chandra ACIS-S \citep{2002PASP..114....1W} observations of TXS\,0128+554 were performed in four separate pointings during March-April 2019 (see Table~\ref{t:chandraobstable} for details). The source was placed at the default aim point on the back-illuminated ACIS charge coupled device (CCD, ACIS-S3). The VFAINT mode and 1/8 CCD readout mode was used to avoid a potential pile-up (see {\it Chandra} Proposer Observatory Guide\footnote{see \url{https://cxc.harvard.edu/proposer/POG/}}). The source was detected in each individual observation, resulting in a total exposure time of 19.3\,ks.

We performed the X-ray analysis of the \chandra observations with CIAO software version 4.12 \citep{2006SPIE.6270E..1VF} and the \chandra calibration data base version 4.9. We reprocessed the pipeline data by running the {\tt chandra\_repro} CIAO tool in order to apply the version 4.9 of the ACIS-S calibration, and used the sub-pixel event repositioning algorithm\footnote{\url{https://cxc.harvard.edu/ciao/why/acissubpix.html}} ({\tt pix\_adj=EDSER}) to obtain the highest resolution X-ray image. 

 We performed the initial inspection of the data in {\tt ds9} \citep{2003ASPC..295..489J} displaying the event files and confirming a strong detection of the source in each individual observation. We defined the source region to be a circle with the 1.5\,arcsec radius corresponding to $\sim 95$\% fraction of the \chandra point spread function, PSF) centered at the source position J2000(RA,Dec) =  (01:31:13.8,+55:45:13.2). We assumed a background region to be an annulus centered on the source position with the radii between 2.5\,arcsec and 4.5\,arcsec. We discuss the details of the X-ray spectral fitting analysis in Section~\ref{xraydatareduction}.

\section{Data analysis}\label{analysis}

\subsection{Radio morphology}

\begin{figure}
\centering
\includegraphics[width=\linewidth]{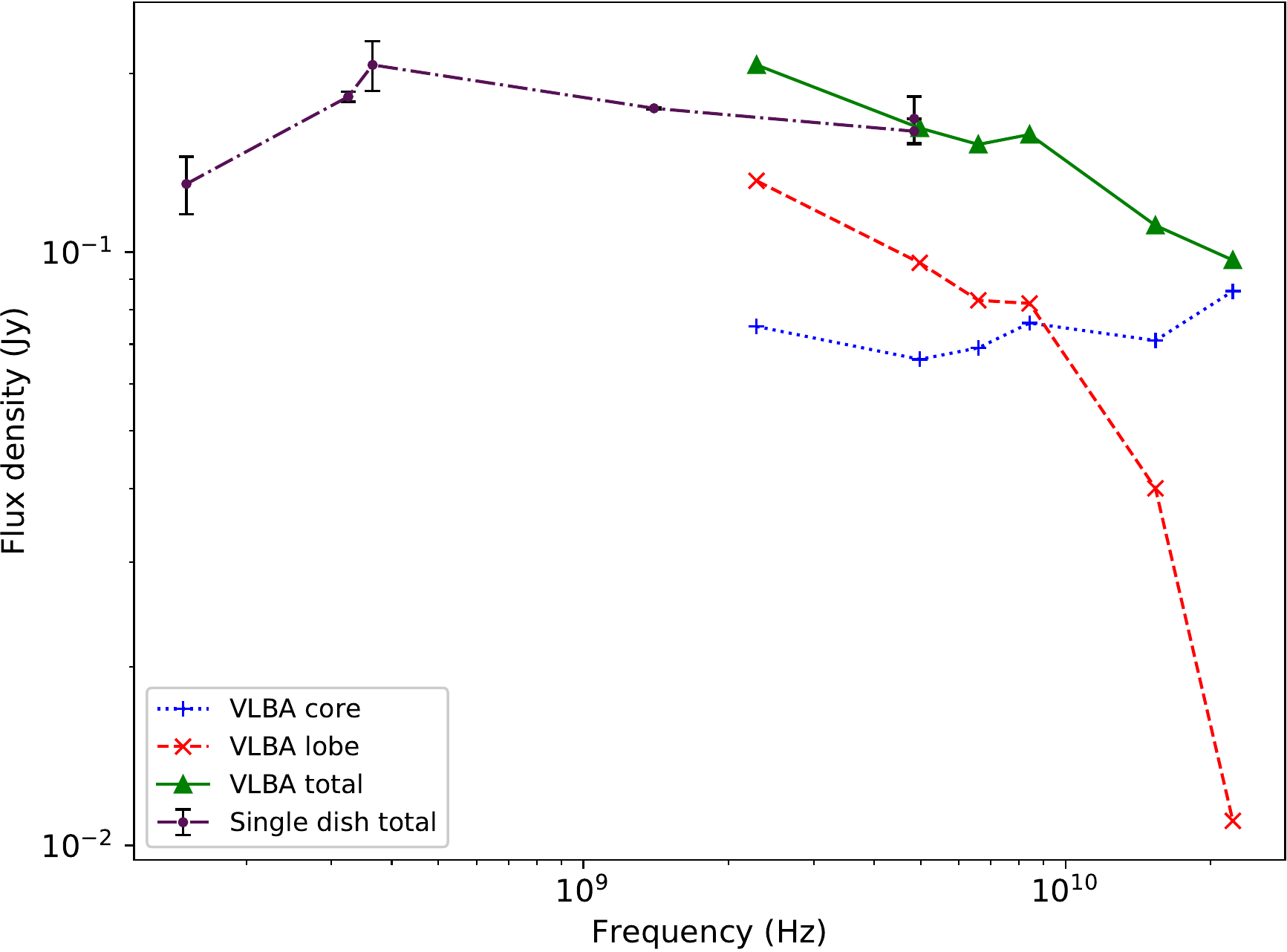}
\caption{\label{VLBA_flux-density} Continuum radio spectrum plot for TXS~0128+554. The single dish flux densities (purple dot-dashed curve) are taken from published surveys as listed in NED. The solid green curve represents total VLBA flux density measurements from our observations, while the red dashed and blue dotted curves represent the pc-scale lobe and core components, respectively.
}
\end{figure}

The radio source TXS~0128+554 is unresolved in the 1.4 GHz NVSS \citep{NVSS},  GMRT \citep{2017AA...598A..78I}, and 2--4 GHz VLASS \citep{VLASS} sky surveys, and in a 5 GHz VLA snapshot image taken in 1994 \citep{1997AAS..122..235L}. The overall radio spectrum is peaked below 1 GHz (\S~\ref{VLBA_flux-density}), with no upturn at low-frequencies that might be expected if significant steep-spectrum extended lobe emission were present. 

The pc-scale radio morphology of TXS~0128+554 (Figure
\ref{multifreqmap}) consists of a bright unresolved core feature
flanked by two arc-shaped radio lobes. This is typical of the
compact symmetric object class of young radio sources \citep{1994ApJ...432L..87W}. The leading edges of the lobes appear brightened,
indicative of interaction with the host galaxy's interstellar medium. The
radio structure is similar to the two-sided inner lobe morphology of
the radio galaxy 3C 84 (NGC 1275; d = 70 Mpc), which has a similar overall diameter ($\sim$ 9 pc) and has been modeled as an expanding plasma cocoon by \cite{2016MNRAS.455.2289F}.

\begin{figure*}
\centering
\includegraphics[width=0.8\linewidth]{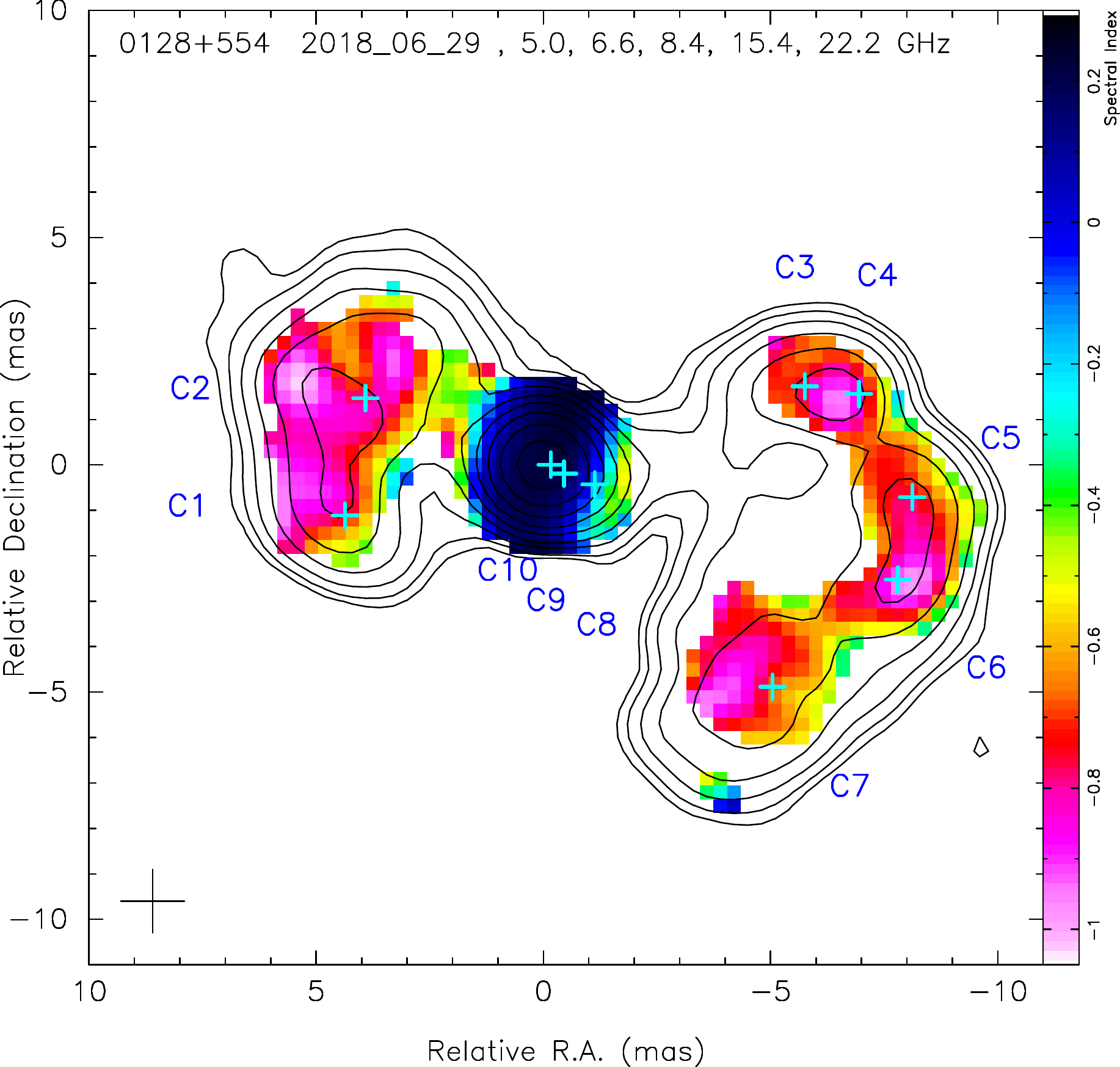}
\caption{\label{specindexC-K}
Total intensity 5 GHz contour map of  TXS~0128$+$554, with spectral index map superimposed in false color. The crosses indicate the Gaussian feature positions fitted to the 15 GHz 2018 June 29 epoch. The spectral indices are derived from a single power-law fit to the pixel values  at 5 GHz, 6.6 GHz, 8.4 GHz, 15.4 GHz, and 22.2 GHz. We aligned the individual frequency maps on the core feature C0 (position at the origin) and restored them with a common circular beam of FWHM = 1.4 mas and a pixel size of 0.3 mas. }
\end{figure*}

The overall projected extent of the lobes is similar in the images at 6.6 GHz and at higher frequencies, ranging from 3.2 pc to 4.3 pc from the core for the eastern lobe, and from 4.5 pc to 6.5 pc for the western lobe.  The 2.3 and 5 GHz images reveal faint steep-spectrum emission regions farther out, at 8.8 pc  (eastern lobe), and at 7.8 pc (western lobe) from the core, respectively. These steep-spectrum regions are aligned along the source axis at $256^\circ$ that is defined by the position angles of features C2 and C6, and the bright inner jet features C8, C9, and C10 (see Fig.~\ref{specindexC-K} and Table~\ref{t:gaussiantable}). 

\subsection{Spectral index map and the core-shift effect}

We created a spectral index map from the 5~GHz -- 22.2~GHz data by aligning the maps on the core feature positions. We checked for possible opacity-related core shifts \citep[e.g.,][]{2011AA...532A..38S,MOJAVE_IX,2018ARep...62..787V} by applying small shifts in different directions and calculating the standard deviation of the spectral index values for all pixels within 1.5 beamwidths of the core position. For all of the frequency maps the minimum standard deviation value occurred for shifts smaller than 0.03 mas, so we chose to not apply any shifts before combining the final maps. A similar approach was used and proven to be efficient and robust by \cite{2019MNRAS.485.1822P}.

We also checked the results of the alignment using an alternative approach to estimate the core shift effect. The algorithm is based on the fast normalized cross-correlation \citep{2D_cross_corr} to register the total intensity maps at different frequencies assuming no strong spectral gradients within the matched areas. We first convolved all the maps with the same circular beam size 1.4~mas and a small pixel size of 0.06~mas, which corresponds to the positional accuracy of the core feature. We then performed the image registration using the optically thin regions of western lobe. We detected no shift for any of the frequency pairs, confirming our other result that was based on minimizing the standard deviation of the spectral index.

To construct the spectral index map we applied a Gaussian taper to the maps above 8 GHz, and restored each map with a circular beam of FWHM = 1.4 mas and a pixel size of 0.3 mas. The former corresponds to the minimum dimension of the uniformly weighted 5 GHz restoring beam. We blanked any pixels with intensity below three times the noise level at any frequency, and calculated the spectral index for the remaining pixels using single power law fits. We note that at the highest frequencies, there is sparser coverage at short spacings in the {\it (u,v)} plane, which can result in less sensitivity to diffuse lobe emission and some artificial steepening of the spectra. 

In Figure~\ref{specindexC-K} we show the spectral index map in false color, superimposed on the 5 GHz total intensity contours. The map reveals an inverted spectrum core region with $\alpha \simeq +0.2$, which steepens to $\alpha \simeq -0.5$ at $r = 2$ mas west of the core. The lobe regions have spectral indices in the range  $-0.7 < \alpha < -1.1$.  The two steepest spectrum regions in the lobes are located on the source axis, slightly further from the core than features C2 and C6. 

Although TXS~0128+554 displays a flat-spectrum core region typically
seen in compact jetted AGN, its negligible core shift is unusual.  For
a typical AGN jet with dominating synchrotron opacity and seen at
small viewing angle, one would expect a core shift between 5 and
22~GHz to be about 0.3~mas (e.g., \citealt{MOJAVE_IX}). This is an
order of magnitude larger than our upper limit that corresponds to a
projected distance of 0.02~pc.  It could result from not being able to
separately resolve the emission from the approaching and receding jets
very close to the core, but we note that there are other high viewing
angle AGN jets with measurable core shifts (e.g., NGC~4261;
\citealt{2015ApJ...807...15H} and NGC~315;
\citealt{2019MNRAS.488..939P}).

Alternatively, we propose that the nature of the apparent radio core
in TXS~0128+554 is not the optically thick $\tau=1$ surface, but
rather a recollimation or reconfinement shock
\citep[e.g.,][]{2014ApJ...787..151C,2013AA...557A.105F,Kovalev_geometry2020},
the position of which is not frequency dependent.  We note that the
inverted spectral index value of $\sim +0.2$ in the core region
further supports this scenario.  The latter could be a result of
absorption by a circumnuclear torus, but additional VLBI spectral line
studies would be needed to further investigate this possibility.

\subsection{\label{kinematics}Multi-epoch VLBA data and jet kinematics}

\begin{figure*}
\centering
\includegraphics[width=0.8\linewidth, trim=0cm 0cm 0cm -0.2cm]{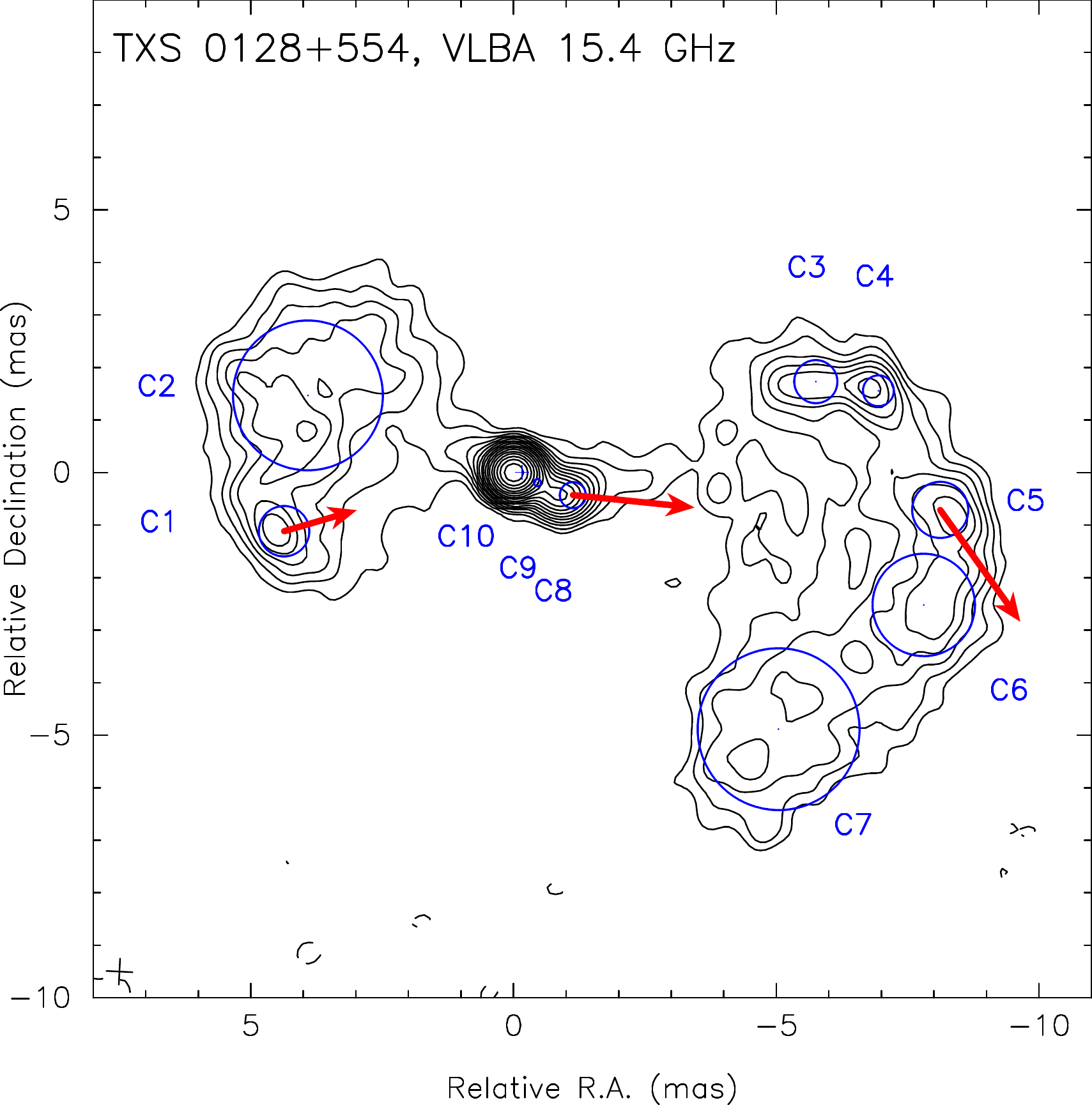}
\caption{\label{stackedimage} Stacked epoch 15 GHz VLBA
  contour image of TXS~0128+554, which combines 8 epochs between 2016 September 26 and 2019 August 23, aligned on the core feature C0 (positioned at the origin) and restored with a circular beam of FWHM = 0.5 mas. The contours are $\sqrt{2}$ times the base contour level of 0.15 mJy per beam. 
  The blue circles indicate the FWHM Gaussian diameters of features
  fit to the 2018 June 29 epoch at 15 GHz, and the red arrows are representative
  of their fitted proper motion vectors. Their lengths correspond to the projected displacement in 20 years assuming constant speed. Only features C1, C8 and C5 have significant proper motion.}
\end{figure*}

TXS~0128+554 was added to the MOJAVE VLBA monitoring program
\citep{MOJAVE_XVII} in 2016 after its detection as a hard-spectrum
(3FHL catalog photon index $\Gamma = 2.1$) \gr source by the {\it Fermi} LAT instrument
\citep{3FHL}. We obtained a total of seven VLBA 15 GHz epochs between
2016 September 26 and 2019 August 23.  In Figure~\ref{stackedimage} we show a stacked total intensity image of these epochs (plus the 2018 June 29 BL251 non-MOJAVE epoch), restored with a circular beam of FWHM = 0.5 mas and pixel size of 0.1 mas. The MOJAVE observations were also sensitive to linear polarization, but we detected no significant polarized flux density at any epoch, with upper limits on the fractional polarization of 0.5\%. 

\begin{deluxetable*}{crrrrrrrrrr} 
\tablewidth{0pt}  
\tablecaption{\label{t:vectormotiontable}Vector Motion Fit Properties of Jet Features}  
\tablehead{ &  \colhead{$\langle S\rangle$}  &\colhead{$\langle R\rangle$} &\colhead{$\langle d_{\mathrm{proj}}\rangle$} & \colhead{$\langle\vartheta\rangle$} & 
 \colhead{$\phi$}&   \colhead{$ |\langle\vartheta\rangle - \phi|$}  &\colhead{$\mu$}  & \colhead{$\beta_{app}$} &  \colhead{$\alpha_m$}& \colhead{$\delta_m$}    \\[-2ex]  
\colhead {I.D.} & \colhead{(mJy)} &\colhead{(mas)} & \colhead{(pc)} & \colhead{(deg)}   & 
\colhead{(deg)}& \colhead{(deg)} &\colhead{($\mu$as y$^{-1})$}& \colhead{($c$)}& \colhead{($\mu$as)}& \colhead{($\mu$as)}
}
\decimalcolnumbers
\startdata 
 C1  & 3 &4.58&  3.28& $ 104.2$ & 286$\pm$20\tablenotemark{a} & 178$\pm$20 & 72$\pm$18 &  0.174$\pm$0.042 &4408$\pm$18 & $-$1111$\pm$28\\ 
 C2  & 15 &4.16&  2.98& $ 71.3$ & 40$\pm$34 & 31$\pm$34 & 59$\pm$31 &  0.143$\pm$0.075 &3963$\pm$19 & 1355$\pm$42\\ 
 C4  & 4 &6.99&  5.00& $ 283.3$ & 257$\pm$28 & 26$\pm$28 & 80$\pm$31 &  0.193$\pm$0.076 &$-$6847$\pm$33 & 1598$\pm$41\\ 
 C5  & 5 &8.16&  5.84& $ 265.1$ & 216$\pm$13 & 50$\pm$13\tablenotemark{b} & 131$\pm$30 &  0.317$\pm$0.072 &$-$8168$\pm$26 & $-$750$\pm$35\\ 
 C6  & 6 &8.26&  5.91& $ 252.4$ & 225$\pm$40 & 28$\pm$39 & 87$\pm$51 &  0.21$\pm$0.12 &$-$7910$\pm$61 & $-$2534$\pm$51\\ 
 C7  & 9 &6.88&  4.92& $ 227.5$ & 63$\pm$85 & 164$\pm$85 & 38$\pm$67 &  0.09$\pm$0.16 &$-$5048$\pm$74 & $-$4639$\pm$72\\ 
 C8  & 6 &1.01&  0.72& $ 245.7$ & 264.1$\pm$4.7 & 18.4$\pm$4.7\tablenotemark{b} & 116$\pm$25 &  0.282$\pm$0.062 &$-$980$\pm$27 & $-$420.8$\pm$9.0\\ 
\enddata 
\tablenotetext{a}{Feature has significant inward motion.}
\tablenotetext{b}{Feature has significant non-radial motion.}
\tablecomments{Columns are as follows:  (1) feature number, (2) mean flux density at 15 GHz in mJy,  (3) mean distance from core feature in mas, (4) mean projected distance from core feature in pc, (5) mean position angle with respect to the core feature in degrees, (6) position angle of velocity vector in degrees, (7) offset between mean position angle and velocity vector position angle in degrees,  (8) proper motion in $\mu$as y$^{-1}$, (9) apparent speed in units of the speed of light, (10) fitted right ascension position with respect to the core at the middle epoch (2018.19) in $\mu$as, (11) fitted declination  position with respect to the core at the middle epoch (2018.19)in $\mu$as.}
\end{deluxetable*} 

We fitted the interferometric visibilities at each 15 GHz epoch using Gaussian features as in \cite{MOJAVE_XVII} and list their properties in Table~\ref{t:gaussiantable}.  The outer jet features (C1 -- C7) display very little flux density variability, and are weak and diffuse, with brightness temperatures ($T_\mathrm{b}$) between $10^6$~K and $10^8$~K. The innermost jet features (C8 -- C10) vary in flux density by a factor of 2--3 over the epochs and are more compact, with $T_\mathrm{b}$ values between $10^8$~K and $10^9$~K.  The core feature maintains a steady brightness temperature of $\sim 10^{10}$~K but shows a drop in flux density by roughly 30\% between 2017 November and 2019 August.  The core $T_\mathrm{b}$ is one to two orders of magnitude lower than typical blazars in the MOJAVE program \citep{2011ApJ...742...27L}, and is similar to those of less Doppler-boosted radio galaxies and members of the CSO/GPS class \citep{2005AA...435..521N}. Stable radio flux densities are a common characteristic of the CSO/GPS class, since their jet emission is less affected by relativistic Doppler boosting \citep{2001AJ....122.1661F}. 
We fit the Gaussian component positions with respect to the (presumed
stationary) core feature at each 15 GHz epoch with a simple vector
motion model, as described in \cite{MOJAVE_XVII}. Our code fits a
constant velocity independently in the right ascension and declination
directions, and solves for the feature position at a reference epoch
date (2018.19) that is midway between the first and last epoch.  We
list the kinematic fit parameters for each feature in
Table~\ref{t:vectormotiontable}. 

Most of the features display no significant proper motion over the
$\sim 3$ year monitoring period, with the exception of C1, C5 and C8. The inner features C9 and C10 do not have sufficient epochs for a reliable proper motion determination, and any apparent motion of the off-axis features may be affected by the three-dimensional expansion of the lobes.
The velocity vector directions of C5 and C8 do not point back to the
direction of the core feature, indicating a current or past
acceleration (Fig.~\ref{stackedimage}). Also, the separation of C1 and
the core is decreasing at a rate of $72 \pm 18$ \muasyr, which may be
due to projection effects, backflow of plasma from the lobe,  or changes in the feature's internal
brightness distribution. A comparison of the 2019 June epoch at 8 GHz with an archival 2010 August epoch (VLBA obscode = BC191I, PI = Jim Condon) shows an outward shift of 127 $\muasyr$ of C5 along PA = $209 \arcdeg$, which is consistent with the C5 speed measurement at 15 GHz from 2016--2019. 

The apparent speeds of C5 (0.32 c $\pm \, 0.07$ c) and C8 (0.28 c $\pm \,0.06$ c) are self-consistent to within the errors, and are similar to the sub-luminal expansion speeds measured in other GPS/CSO jets \citep{2005ApJ...622..136G}. Given the double-lobed morphology of this source, its jets likely lie close to the plane of the sky (see \S~\ref{intrinsic}), and superluminal motion is therefore not expected.

\subsection{\label{intrinsic}Intrinsic jet speed and viewing angle}

The asymmetry of the jet morphology and reasonably high apparent velocities of features C5 and C8 suggest that the radio properties of TXS~0128+554 are affected by relativistic beaming and time delay effects. In this case, the western (approaching) jet and lobe emission are Doppler boosted, while the eastern lobe emission is de-boosted.  The eastern lobe emission also has to travel a longer distance to the observer, so we are seeing it at an earlier stage of its evolution (and thus with a smaller apparent size). In the same vein, the photons we see from the western lobe were emitted at a later stage of the sources' evolution than those from the core region. 

Under the assumption of identical jet and counter-jet expansion speeds $v = \beta c$ at viewing angles $\theta$ and $180^\circ - \theta$ from the line of sight, respectively, we can use the apparent separation of two jet features located on opposite sides of the core to measure $\beta\cos{\theta}$ \citep{1967MNRAS.136..123R}.  Considering only time-lag effects, the expected separation ratio is 
$$Q = {1 + \beta\cos{\theta} \over 1 - \beta\cos{\theta}}.$$

Because the location of C2 represents the centroid of a complex emission region, we use the two steepest spectrum locations in the lobes, at $r = 5.6 \pm 0.1$ mas and $r = 8.4 \pm 0.1$ mas along the main jet axis (PA = $256^\circ$). This yields $Q = 1.50 \pm 0.03$ and $\beta\cos{\theta} = 0.20 \pm 0.01$. The separation ratio for the mean positions of features C4 and C1 (Table~\ref{t:vectormotiontable}) yields a similar value of $Q = 1.53 \pm 0.03$, which would be expected if the lobes are expanding radially from the central source. 

The expected flux density ratio due to relativistic boosting is
$$J = \left[{1 + \beta\cos{\theta} \over 1 - \beta\cos{\theta}}\right]^p.$$
The boosting index $p$ is $2-\alpha$ for continuous jet emission, and $3-\alpha$ for an isolated bright feature, where $\alpha$ is the spectral index \citep{UP95}. 

There are several possible ways to measure $J$ in the source. The first is to sum the emission from each lobe, excluding the core feature, which gives $J = (42 \pm 2 \;\mathrm{mJy}) / (21 \pm 1 \;\mathrm{mJy}) = 2.0 \pm 0.1$. A second method involves taking a slice along the main jet axis in the 15 GHz stacked epoch image. There is an initial peak of $J = 5.5 \pm 0.5$ at the location of feature C8, where $\alpha = -0.07 \pm 0.04$. $J$ drops below unity until $r = 6.2$ mas, and then rises to a plateau of $J = 1.8 \pm 0.1$ at $r = 6.5$ mas. Past $r \gtrsim 7$ mas, $J$ rapidly rises above 8 due to the smaller size of the eastern lobe. 

The interpretation of the jet/counter-jet flux density ratio is complicated by the time delay effects in the two lobes. For example, if feature C8 represents a (more recent) large jet outburst, its counterpart feature in the eastern jet may be located very close to the core and not currently resolvable in the 15 GHz image. Likewise, the total flux density of each lobe represents emission at different evolutionary stages and thus may not be intrinsically identical. 

\begin{figure}
\centering
\includegraphics[width=\linewidth]{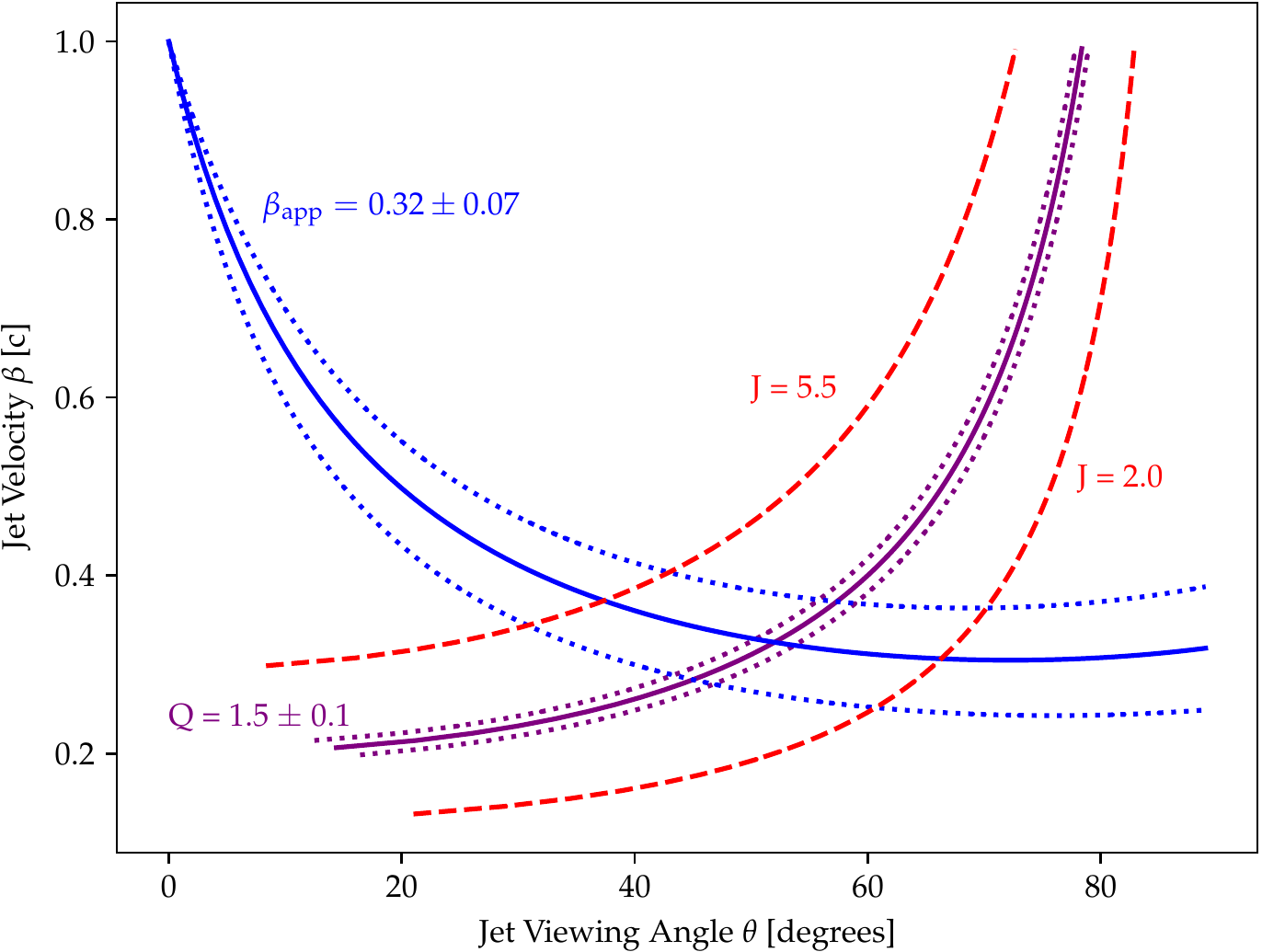}
\caption{\label{betacostheta}
Diagram showing constraints on jet speed $\beta$ and viewing angle $\theta$ derived from the apparent jet speed (blue solid curve), jet/counter-jet length ratio (purple solid curve) and jet/counter-jet flux density ratios (red dashed curves). The dotted curves represent the $\pm 1 \sigma$ ranges on the measured quantities.  The intersection point of the $\beta_\mathrm{app}$ and $Q$ curves is at $\theta = 52^\circ, \beta = 0.32$.
}
\end{figure}

In Figure~\ref{betacostheta} we plot the constraints on $\beta\cos{\theta}$ based on the measured apparent speed, separation ratio, and flux density ratio. For the latter, we assume $p = 2.8$ and plot two dashed curves, for $J = 2.0$ and $J = 5.5$. The dotted curves represent the $\pm 1 \sigma$ ranges. The measured quantities constrain the intrinsic jet velocity to the range $0.25 < \beta < 0.4$. The apparent speed and jet length ratio provide a narrower constraint on the viewing angle ($43^\circ < \theta < 59^\circ$) than the flux density ratios ($31^\circ < \theta < 70^\circ$).  The intersection point of the $\beta_\mathrm{app}$ and $Q$ curves is at $\theta = 52^\circ, \beta = 0.32$, which corresponds to mild Doppler boosting factors of $\delta = 1.2$ in the approaching jet and $\delta = 0.79$ in the counter-jet. 

By assuming that the plasma in the western lobe has been expanding at constant velocity of $0.32 \; c$ and using its de-projected size, we derive a kinematic age of $r_\mathrm{max} / c\beta_\mathrm{app} = 82 \pm 17$ years. This places TXS~0128+554 among the youngest known relativistic AGN jet systems. We compare its properties to other recently launched AGN jets in Section~\ref{agn_comparisons}. 

\subsection{\label{xraydatareduction}Chandra X-ray Analysis}

\begin{figure}
\centering
\includegraphics[width=\linewidth]{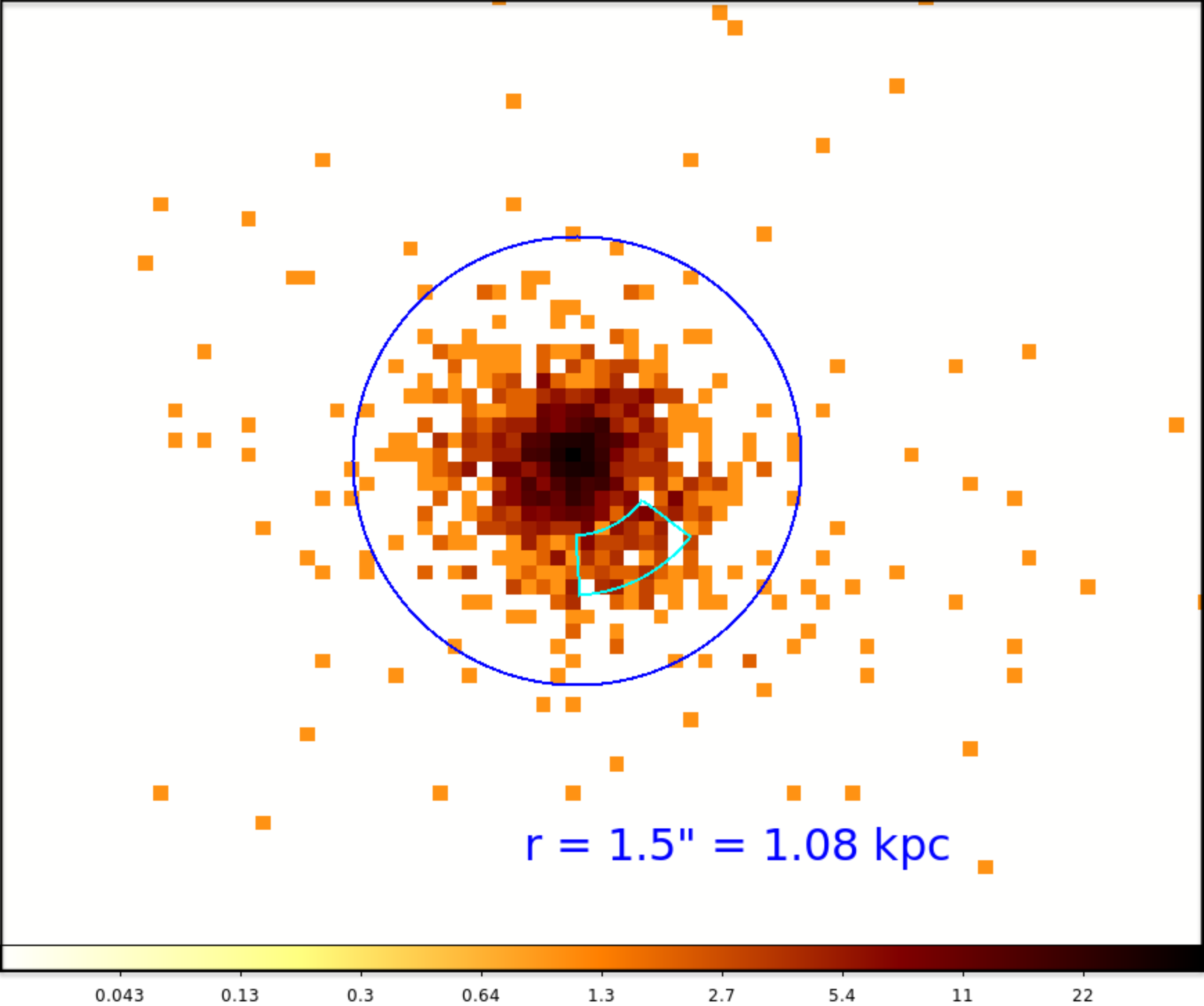}
\caption{\label{xrayimage} The {\it Chandra} ACIS-S X-ray image centered on TXS~0128+554 (image size 7.5\arcsec$\times$6.2\arcsec). The four individual observations have been merged. The image is binned to 0.246\arcsec, half of the native ACIS pixel size. The circle marks the region assumed for the spectral extraction (radius of 1.5\arcsec = 1.08\,kpc) and containing 95\% fraction of the PSF counts.
The {\it Chandra} PSF artifact region is marked with the cyan color. The color intensity represents the number of counts per pixel with the color bar showing the scale.}
\end{figure}

The {\it Chandra} ACIS-S observations of TXS\,0128+554 were taken at four separate epochs and needed to be merged into one event file for detailed image analysis of the spatial properties of the source, and to look for any extended structures.
We merged individual {\it Chandra} observations using {\tt reproject\_obs} in CIAO. For each observation we simulated the {\it Chandra}
point spread function (PSF) using {\tt simulate\_psf}, and merged the individual simulations into one image file. We then used both images to generate the radial surface brightness profiles of the source and the PSF, assuming a set of annuli centered at the (J2000) coordinates: RA=1:31:13.8588, DEC=+55:45:13.532. We excluded the position angle covered by the known PSF asymmetry region given by {\tt make\_psf\_asymmetry\_region}. The final radial profiles were fit to check for any extended features and were found to be consistent with a point source. 
We show the merged {\it Chandra} image centered on TXS\,0128+554 in Fig.~\ref{xrayimage}.

We extracted the spectra and all the corresponding  standard calibration files (i.e., {\tt arf} and {\tt rmf}) for each observation using {\tt specextract} in CIAO\footnote{\url{https://cxc.harvard.edu/ciao/}}, assuming the source and background regions defined above.
We performed all spectral modeling of the {\it Chandra} X-ray data in {\it Sherpa}\footnote{\url{http://cxc.harvard.edu/sherpa/}}
\citep{2001SPIE.4477...76F,2009pysc.conf...51R}.  We used Cash and Cstat fitting statistics \citep{1979ApJ...228..939C} and the Nelder-Mead optimization method \citep{10.1093/comjnl/7.4.308}.
We tabulate the fit parameters resulting from a simultaneous fit of all spectra in Table~\ref{t:models}.  The fitted total X-ray luminosities are  $3.2\times10^{42}\, \rm erg\,s^{-1}$ between 0.5 keV and 2 keV, and $2.1\times10^{42}\, \rm erg \, s^{-1}$ between 2 keV and 10 keV. 

\begin{deluxetable*}{cllcccc} 
    \tablecaption{\label{t:models} X-ray Spectral Models}
    \tablehead{ \colhead{Model}  & \colhead{ N$_H$(z)$^b$ } & \colhead{ $\Gamma$ } 
    & \colhead{kT} & \colhead{Flux} & \colhead{Flux} & \colhead{Cash/d.o.f} \\[-2ex]
    description$^a$ & \colhead{(10$^{22}$ cm$^{-2}$)}& & \colhead{(keV)} 
    & \colhead{(soft)}& \colhead{(hard)}& }
\startdata
PL & $0.70\pm0.08$ & $2.38\pm0.10$ & & $10.8\pm1.31$ & $7.04^{+1.45}_{-1.23}$ & 1669.9/1773\\
PL + Apec & 0.67$^{+0.08}_{-0.05}$ & 2.39$^{+0.06}_{-0.11}$ & $<0.08$ & $11.2\pm0.22$ & $7.03^{+0.31}_{-0.25}$ & 1668.8/1771 \\
\enddata

\tablenotetext{a}{ PL - power law, Apec - thermal hot plasma model, G - Gaussian line;  
Unabsorbed flux in the soft (0.5-2\,keV) and hard (2-10\,keV) bands extrapolated 
from the model fitted over the 0.5-7\,kV range, in units of 10$^{-13}$\,erg\,cm$^{-2}$\, s$^{-1}$.}

\tablenotetext{b}{All models assume the Galactic absorbing column of $2.633\times 10^{20}$\,cm$^{-2}$ and
redshift $z = 0.0365$ for the intrinsic absorption.}

\tablenotetext{c}{Models: {\tt 1/ phabs*zphabs*powerlaw;
2/ phabs*zphabs*(powerlaw + apec).
}}

\end{deluxetable*} 

In all models we assumed a constant Galactic absorption column of $2.6\times10^{20}$ cm$^{-2}$ (from COLDEN\footnote{\url{https://cxc.harvard.edu/toolkit/colden.jsp}}, \citealt{1992ApJS...79...77S}). We initially fit an absorbed power law model and obtained the best fit photon index, $\Gamma = 2.38\pm0.10$ and an intrinsic absorber at $z=0.0365$ with an equivalent column of hydrogen of ${\rm N}_H(z)=6.7\times10^{21}$\,cm$^{-2}$. 
The latter is in agreement with the \HI absorbing column density of $1.2 \times 10^{21} \rm \,cm^{-2}$ predicted from the $N_{\matHI}$ -- linear size relation found by \cite{2003AA...404..871P}. 

Recent studies of neutral $N_{\matHI}$ (21\,cm) and total $N_{H}$ (X-rays)
absorbing columns in compact radio AGN indicate a correlation between the two column densities, with a scatter related to the spin temperature and a covering fraction of the absorbing gas \citep{2010ApJ...715.1071O,2017ApJ...849...34O}. 
Our X-ray measurement of $N_H(z)$ and the extrapolated $N_{\matHI}$ value place TXS\,0128+544 at the upper part of this scatter, indicating a larger amount of neutral medium in comparison to the sources with similar $N_{H}$. A future detection of 21\,cm absorption could provide an important data point for this correlation. Interestingly, the TXS\,0128+544 X-ray absorption is in agreement with the relation between the radio source size, radio luminosity and a total $N_H$ column studied by \cite{2019ApJ...871...71S}, in which the more compact CSOs show higher absorption columns. However, these correlation results were based on a small number of CSOs observed in X-rays and they require confirmation with larger samples.

The photon index of $\Gamma=2.38$ is steep in comparison to a more typical value of $\Gamma_\mathrm{ave} \sim 1.7$ found in other CSOs with {\it Chandra} spectra \citep{2016ApJ...823...57S}. The soft X-ray spectrum might due to a hot thermal gas in the central kpc-scale region. We therefore added a thermal component to the spectral model and
obtained an upper limit on the temperature of $\rm kT = 0.08\, \rm keV$ (APEC\footnote{\url{https://heasarc.gsfc.nasa.gov/xanadu/xspec/manual/XSmodelApec.html}; \url{http://atomdb.org/}} model in {\it Sherpa}). The contribution from such thermal emission would dominate the spectrum at lower energies. However, the current {\it Chandra} spectra do not provide any constraint on the fraction of thermal radiation within the nuclear, $<1$\,kpc region, and no diffuse X-ray hot gas outside that region above the background flux level of
$2.6\times10^{-15}$\,erg\,s$^{-1}$\,cm$^{-2}$.

We also looked for an iron line at 6.4\,keV often indicating a presence of a reflection of the primary absorbed X-ray radiation off the cold medium (e.g., a large scale clumpy torus; see \citealt{2018ARA&A..56..625H} for a review).
We fit the spectrum with an absorbed power law model (fixed $N_H(z)=7\times10^{21}$\,cm$^{-2}$) and an unresolved narrow Gaussian line (0.1\,keV width).
The line was not detected and we obtained an upper limit on the normalization of the line of $< 1.2\times10^{-6}$\,photons\,s$^{-1}$\,cm$^{-2}$ corresponding to a 3$\sigma$ upper limit of 780\,eV on the line's equivalent width. We ran simulations, following
\cite{2002ApJ...571..545P}, to test for the presence of the iron line, resulting in a $p$-value of 0.832. This indicated that the spectra are consistent with the absorbed power law model and no additional line is required by the current data.
Thus we do not find any strong reflection signatures in the {\it Chandra} spectrum.

Future high resolution, high S/N spectra with addition of the high energy X-rays are needed for any detection of spectral or reflection features and more detailed study of the X-ray nature of this source.  

\section{Discussion}
\subsection{Cocoon model}

The overall compact radio morphology of TXS~0128+554 can be attributed to a highly over-pressurized cocoon structure that is driving a bow shock into the surrounding galactic medium. In this scenario a thin shell exists around the cocoon and is the site of particle acceleration, with synchrotron emission being produced by the non-thermal electrons. Simple analytical models (e.g., \citealt{1989ApJ...345L..21B}, \citealt{2005MNRAS.364..659K}, \citealt{2008ApJ...680..911S}) have been developed 
to describe the evolution of the cocoon as it is inflated by a relativistic jet outflow. These generally contain several features associated with the jet and counter-jet: a head (hotspot) where the collimated outflow impacts the external medium, and a working surface over which the jet kinetic energy is dissipated. The latter is the leading curved edge of a radio lobe embedded within the cocoon (see, e.g., Fig. 1 of \citealt{1989ApJ...345L..21B}, and Fig. 1 of \citealt{2009MNRAS.395.1999C}). The jet head advances at a speed $v_h$ dictated by ram pressure equilibrium, such that $L_j/c = \rho v_h^2 A_h$, where $\rho$ is the external medium density, $L_j$ is the jet kinetic power, $c$ is the speed of light, and $A_h$ is the area of the working surface.

The hotspots of radio galaxies (e.g., Cyg A, see
\citealt{1996AARv...7....1C}) frequently exhibit a flat or inverted
radio spectra ($\alpha > -0.5$), implying that they are principal
sites of particle acceleration \citep{1989AA...219...63M}. Such active
hotspots are not seen in TXS~0128+554. All of the inverted spectrum
regions are located in the bright inner jet region close to the core.
Those regions are also bright in the higher-frequency VLBI bands,
indicating ongoing particle acceleration. In contrast, the radio lobes
of the source have steep-spectra, and their low frequency emission is
likely due to an aging population of electrons.  

The distinct gap between the inner and outer regions in the VLBI
images above 6 GHz leads us to conclude that they are not causally
connected.  The brightness temperature and steep spectral index of C5 are
comparable to the other bright features in the eastern lobe, whereas
its radio brightness might be expected to be enhanced if it is
actively being energized by the C10-C8 jet. Although the velocity
vector of C8 does point to C5, the velocity vectors of these two
features are not collinear, suggesting that their similar apparent
speeds are merely a coincidence.

We conclude that the jets in TXS~0128+554 turned off for a period
during its lifetime, and have subsequently re-started. Such
intermittent jet activity has been proposed in theoretical models for
young radio sources, e.g.,
\citep{1997ApJ...487L.135R,2009ApJ...698..840C}.  The radio lobes are
remnants from the initial period of jet activity, while the inner jet
(associated with core features C8, C9 and C10; \S~\ref{kinematics})
was launched more recently and is advancing during the present epoch.

\begin{figure}
\centering
\includegraphics[width=\linewidth]{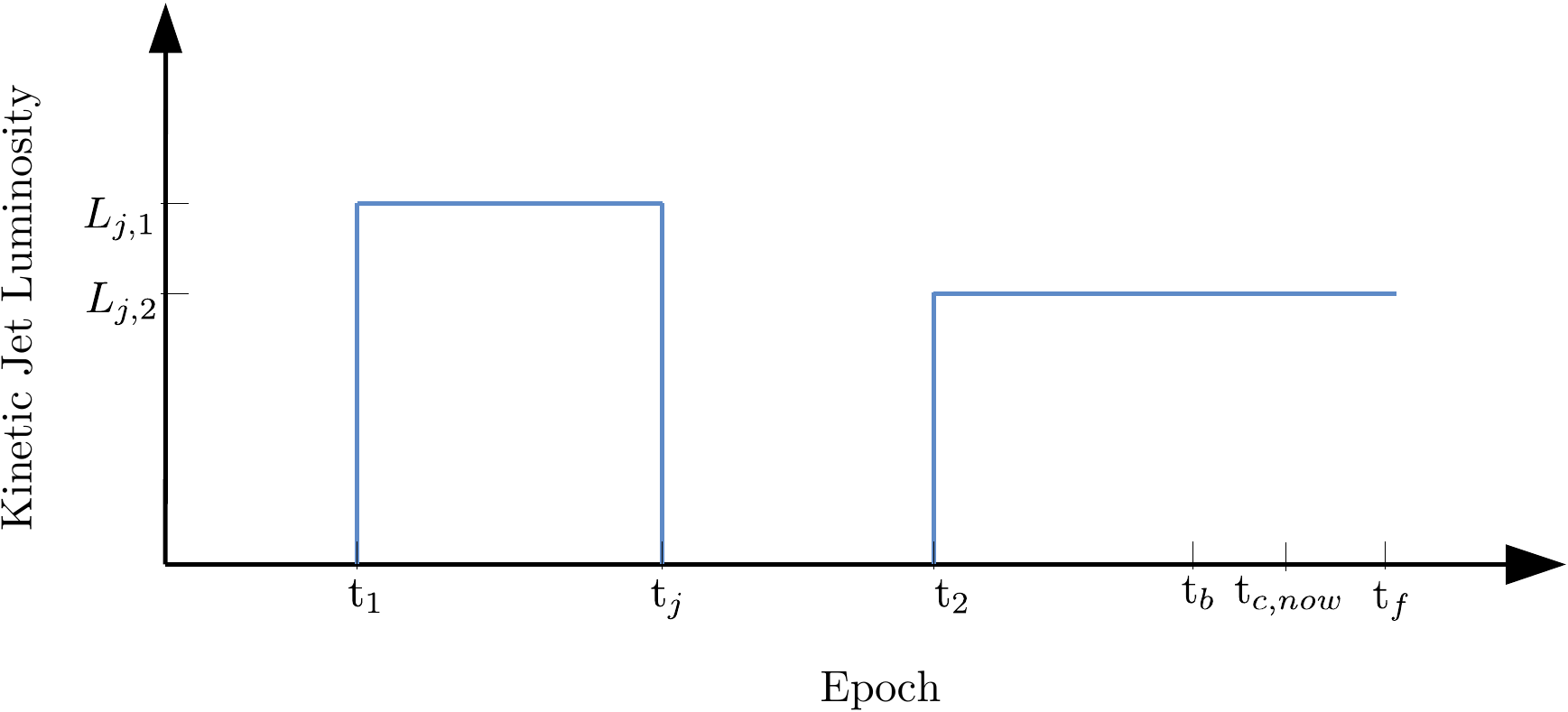}
\caption{\label{jet-timeline} Schematic diagram of the evolution of kinetic jet luminosity in TXS~0128+554 with times in the rest frame of the central engine (in arbitrary units and not to scale). The first jet turn-off happened at a time $t_j$ after its initial birth at $t_1$. A period of jet inactivity lasted until the jet re-started at $t_2$. The latter event generated the bright $\sim 2$ mas long inner jet visible in the VLBA 15.4 GHz image at the current epoch $t_{c,\mathrm{now}}$. Due to light time delays and the relativistic expansion speeds, radiation from the western (approaching) lobe, which originated from the initial jet activity epoch, is seen at a later epoch than the core emission, while the eastern (receding) lobe is viewed at an earlier epoch than the core. }
\end{figure}

In Figure~\ref{jet-timeline}  we show a schematic timeline of the jet activity in the source. The marked epochs refer to times in the rest frame of the central engine. The jet was initially launched at $t_1$, and turned off at $t_j$, producing the radio lobes that we see today. The bright C8 radio feature marks the outer edge of a restarted jet that was launched at $t_2$. Photons we currently receive from the core region were emitted at $t_{c,\mathrm{now}}$. The shorter light travel distance to the western lobe and its relativistic advance speed imply that its photons were emitted at a later time $t_f$. Conversely, we see a relatively younger version of the eastern lobe  (i.e., at epoch $t_b$) since it is farther away and receding with respect to its western counterpart. For a constant advance speed, we would expect \citep{1967MNRAS.136..123R}
$$t_f = t_{c,\mathrm{now}}/(1-\beta\cos{\theta}) = 1.25\, t_{c,\mathrm{now}} $$
and $$ t_b = t_{c,\mathrm{now}}/(1+\beta\cos{\theta}) = 0.84\, t_{c,\mathrm{now}}.$$

Clearly $t_j$ is an important parameter since it (along with $L_{j,1}$) dictates how much total energy has been injected into the lobe structure.  Given the young kinematic age of $82\pm 17$ y, there is insufficient time for the lobes to experience much spectral ageing at low radio frequencies, and we see little indication of curvature in our radio spectra of the bright lobe features. Assuming a magnetic field strength typical of CSOs, a spectral downturn would occur at 22 GHz in the lobes of TXS~0128+554 only after $\sim1700$ y \citep{1970ranp.book.....P}. 

We can place a conservative lower limit on the duration of the initial period of
jet activity by equating the source cavity enthalpy to the energy
injected during the initial epoch of jet activity, i.e.,
$4pV=L_{j,1}(t_j-t_1 )$. If we were to assume that the maximum jet
kinetic luminosity TXS 0128+554 could have had in this phase was
equal to that of the most powerful known CSOs ($L_{j,1} \simeq
10^{45}\, \rm erg\,s^{-1}$; \citealt{2020ApJ...892..116W}), we find
$t_j - t_1 \geq 2.7$ y.

We apply the model of \cite{2008ApJ...680..911S} to the first epoch of jet activity which is responsible for production of the outer radio lobes. Here the jet kinetic power is constant over time, and the external medium has a flat density profile $\rho = m_p n_o$, where $m_p$ is the proton mass and $n_o = 0.1$ cm$^{-3}$ (typical for CSO environments). This leads to a constant head advance speed over time. This is a reasonable assumption as we see no evidence of a significant change in speed in the moving features over the three year period of the MOJAVE 15 GHz VLBA observations, and over a 10 year period at 8 GHz (\S~\ref{kinematics}).

The energy transported by the jet is transformed into the cocoon's internal pressure $p$, which also causes it to   expand sideways with a velocity $v_c = (p/\rho)^{1/2}$. Additionally, the transverse size $l_c$  is taken to vary with time as $l_c\propto t^{1/2}$, so as to reproduce the ballistic phase of jet advance, following the numerical simulations of \cite{2002MNRAS.331..615S}. The model has three free parameters: $v_h$, $l_c$, and the linear size ($LS$), with the latter being defined as the separation of the jet head from the core. From the observations, the parameter values at epoch $t_f$ are $v_h = 0.32 \pm 0.07$ c, $l_c = 2.87 \pm 0.36$ pc, and $LS = 5.67 \pm 0.65$ pc.  This yields a jet kinetic power of $(1.28\pm1.16) \times10^{43}\, \rm erg\,s^{-1}$,  which agrees to within an order of magnitude with the predicted value based on previously published correlations with jet radio luminosity \citep{1999MNRAS.309.1017W, 2012MNRAS.423.2498D}. We note that this is an estimate of the jet kinetic power $L_{j,1}$ of the initial jet activity epoch ($t_1 < t < t_j$) only,  since  this method takes into consideration the features of the lobes only and not the (re-started) inner jet.

The radio emission from the lobes, as in other GPS/CSS sources, is most likely of synchrotron origin.  The lobe is injected with freshly accelerated electrons from the terminal jet hotspot, which then undergo adiabatic and radiative cooling. We assume a power-law energy distribution injection function that stays fairly constant with time during the initial jet activity (ballistic expansion) phase. By further assuming the electron energy density to be in rough equipartition with the magnetic energy density and lobe pressure, the lobe synchrotron luminosity is predicted to be log L=$41\pm1\, \rm erg\,s^{-1}$. This is in agreement with the integrated radio luminosity, $2.8\times10^{41}\, \rm erg\,s^{-1}$,  which we obtained by fitting a single component synchrotron self-absorption (SSA) model to the single dish and VLBI radio flux densities as plotted in Figure 2. The fit yields a SSA turnover frequency of $\sim657\, \rm MHz$, which is in agreement with the $\sim630\, \rm MHz$ turnover that is predicted by the \cite{2008ApJ...680..911S} cocoon model. 

Equipartition magnetic field estimates based on a spherical cocoon geometry \citep[e.g.,][]{2012MNRAS.424..532O, 1970ranp.book.....P} yield values in the range $3\mbox{-}9$ mG. The \cite{2008ApJ...680..911S} cocoon model yields a similar range: $\sim5\mbox{-}20$ mG, depending on the choices for the magnetic field and electron energy equipartition parameters $\eta_B$ and $\eta_e$. This is comparable to the typical range of $1\mbox{-}10$ mG measured in GPS and CSO galaxies (\citealt{2002AA...389..115D}, \citealt{2006AA...450..959O}).

It is somewhat surprising that TXS~0128+554 lies significantly outside
the well-studied trend between turnover frequency and linear size
(e.g., \citealt{ODe97}), which would predict a much higher turnover
frequency of $\sim 50$~GHz. \cite{2008AA...482..483T} found a cluster
of several GPS sources that lie in a similar location in the turnover
-- linear size plane. In the case of TXS~0128+554 this may be due to
the fact that the outer lobes have not been re-supplied by the central
AGN for some time, which has affected their overall evolution and
radio properties.   

With a 150 MHz radio luminosity of only $10^{23.6}$ W Hz$^{-1}$,
TXS~0128+554 is at the low end of the luminosity range of a large
sample of over 23 000 AGN from the LOFAR 1st data release studied by
\cite{2019AA...622A..12H}. These authors found that most AGN in the
luminosity range $10^{25} < L_{150} < 10^{27}$ W Hz$^{-1}$ are
long-lived objects in relatively poor (group-like) environments, and
there are relatively few physically small objects in this luminosity
range. The large number of small objects at lower luminosities led
them to conclude that they must either have a different lifetime
distribution, or different jet physics from the more powerful objects.
It may be possible, therefore, that many weak AGN like TXS~0128+554
never grow to large sizes, due to their episodic/short-lived jet
activity. 

\begin{figure}[t]
\centering
\includegraphics[width=\linewidth]{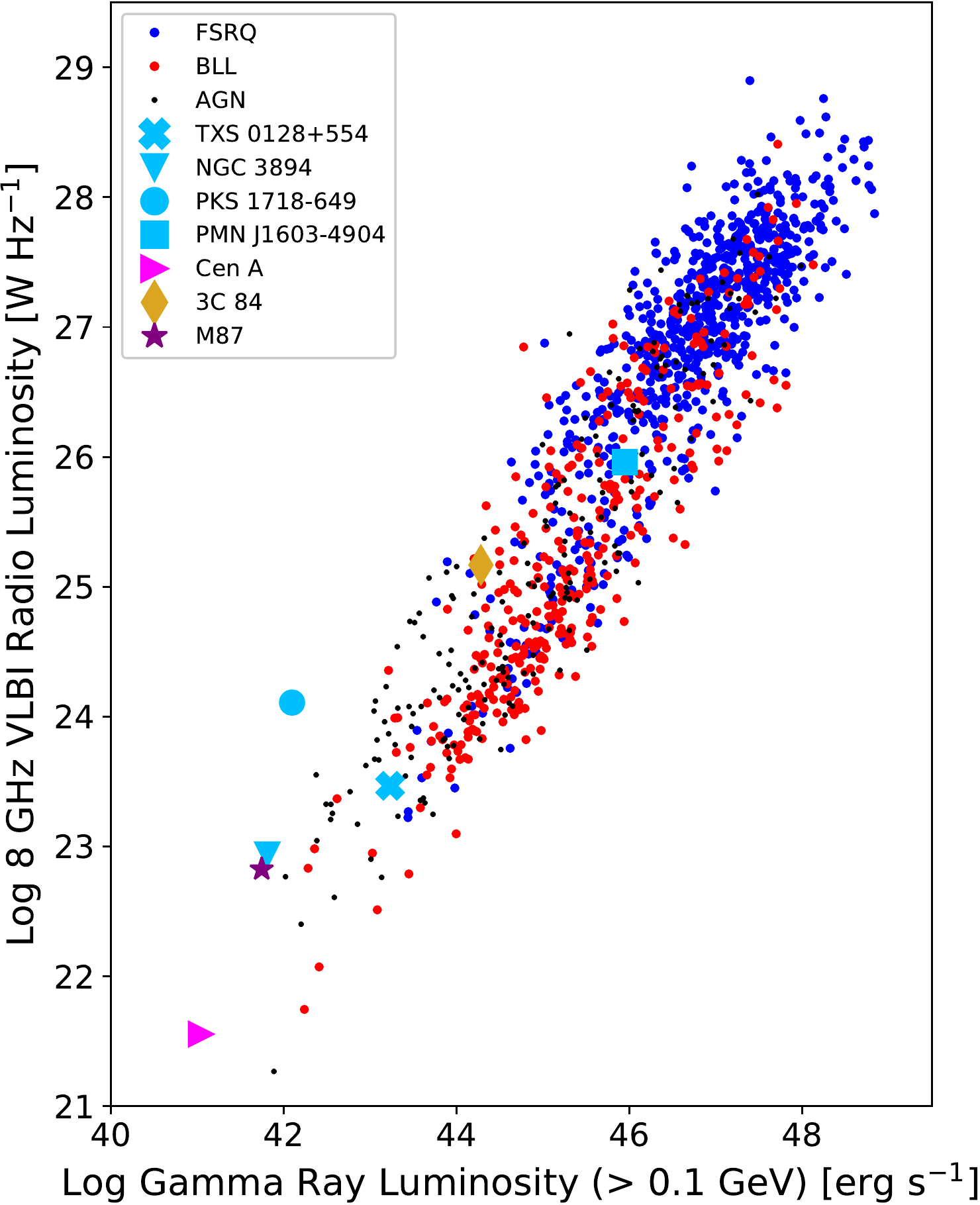}
\caption{\label{vlbi8ghzvsgamraylum} Plot of VLBI 8 GHz luminosity versus \gr luminosity ($> 0.1$ GeV) for AGN in the 4LAC catalog. Small red circles denote BL Lac objects, small blue circles denote flat-spectrum radio quasars, and black dots denote other AGN. The blue cross denotes TXS~0128+554, and large light blue symbols represent \textit{Fermi}-detected CSOs. The other large symbols denote selected nearby radio galaxies detected by {\it Fermi}.}
\end{figure}

\begin{figure}[t]
\centering
\includegraphics[width=\linewidth]{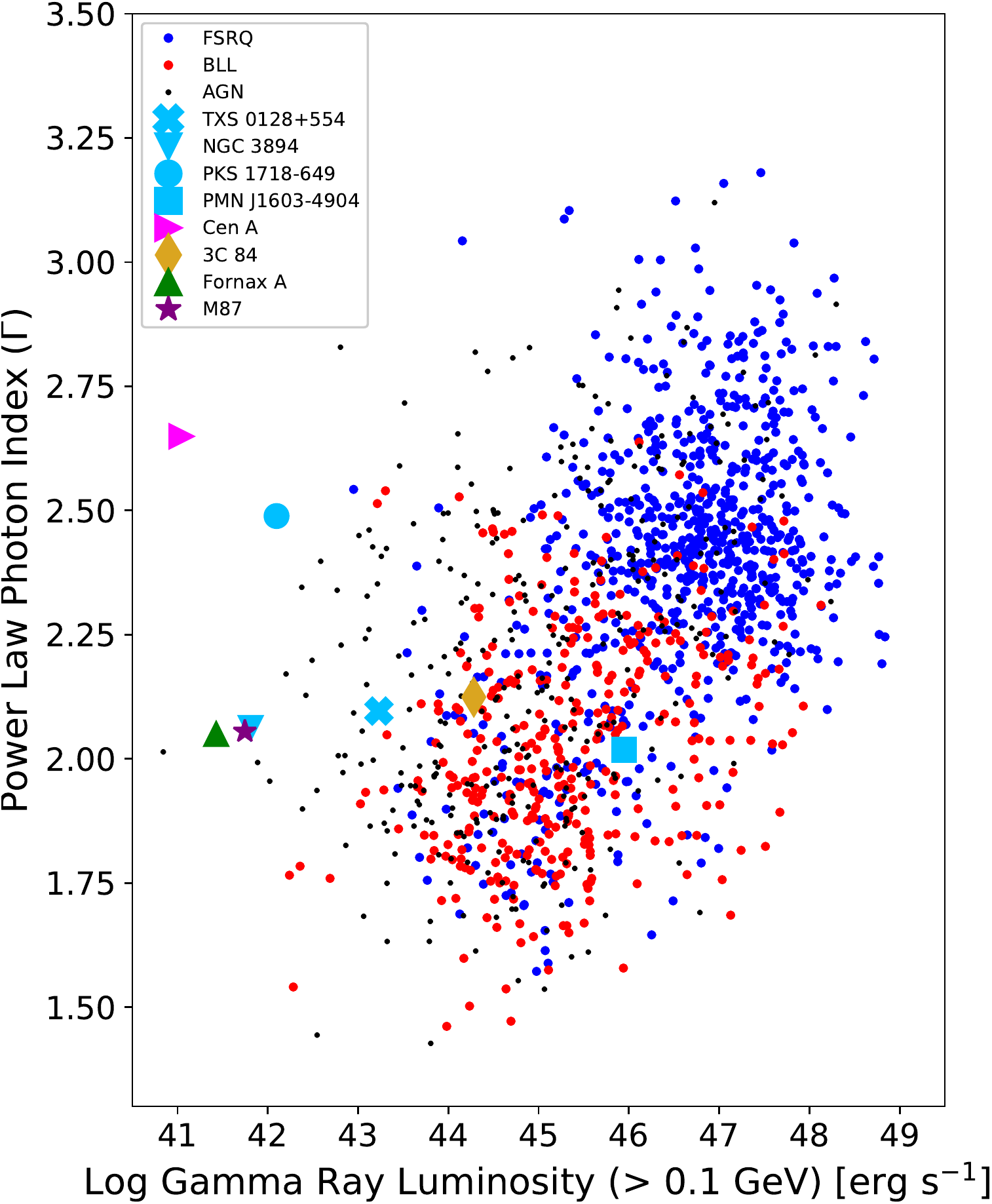}
\caption{\label{plindexvsgamraylum} Plot of power law photon index versus \gr luminosity ($> 0.1$ GeV) for AGN in the 4LAC catalog. Small red circles denote BL Lac objects, small blue circles denote flat-spectrum radio quasars, and black dots denote other AGN. The blue cross denotes TXS~0128+554, and large light blue symbols represent \textit{Fermi}-detected CSOs. The other large symbols denote selected nearby radio galaxies detected by {\it Fermi}.}
\end{figure}

\subsection{Origin of the high-energy emission}

\cite{2008ApJ...680..911S} predict that the lobes in powerful GPS/CSS sources should also emit significant \gr emission via inverse Comptonization of seed photons by the lobe electrons. Stellar emission in the near-IR from the host galaxy, synchrotron emission from the lobe electrons, IR radiation from the dusty torus, UV radiation from the accretion disk and cosmic microwave background radiation are considered to be the main components for the seed photon field. However, when applied to our source, the model under-predicts the \gr flux observed by Fermi-LAT by three orders of magnitude. We therefore conclude that it is highly unlikely that the observed \gr emission of TXS~0128+554 originates in its lobes.  

The other possible sources of X-rays and $\gamma$-rays are the inner jet/optically thick core region and a shocked shell surrounding the cocoon. We can compare the un-absorbed X-ray luminosities of the cores and bright knots or hotspots of some of well-known, {\it Chandra}-resolved jets to that of TXS~0128+554 to examine whether the majority of the high energy emission may originate in the radio-loud inner core. We consider the cases of M87 and Cen A, which are both nearby and well-resolved in X-rays. The core of M87 has a luminosity of $4.4\times10^{40}\, \rm erg\,s^{-1}$ in the 0.5--10 keV band \citep{2002ApJ...564..683M}, while the HST-1 feature reached an order of higher magnitude X-ray luminosity during a large flare \citep{2006ApJ...640..211H}. 
Cen A is slightly more luminous in X-rays; the nucleus has a luminosity of $5\times10^{41}\, \rm erg\,s^{-1}$ in the 2--10 keV band \citep{2000ApJ...531L...9K} while the brighter knots have luminosities of the order of $10^{38}$--$10^{39}\, \rm erg\,s^{-1}$ in the 0.1--10 keV band \citep{2001ApJ...560..675K}. Hence, when compared to the unbeamed X-ray luminosity of TXS~0128+554 ($2.30\times10^{42}\, \rm erg\,s^{-1}$ and $1.51\times10^{42}\, \rm erg\,s^{-1}$ in the 0.5--2 keV and 2--10 keV bands, respectively), it is plausible that the core is the major source of the high energy emission. 

We note that the shell surrounding the south-west lobe of Cen A has also been observed to emit strongly in X-rays ($\sim10^{38}\, \rm erg\,s^{-1}$), and the majority of the X-ray emission has been shown to be non-thermal \citep{2009MNRAS.395.1999C}. The jet power of Cen A, calculated from the size of the cavity excavated by the jets, is found to be $\sim10^{43}\, \rm erg\,s^{-1}$. This is similar to that of TXS~0128+554, although Cen A is considerably older, with a kinematic age of $\sim2\times10^6\, \rm years$. 
 
\subsection{\label{agn_comparisons} Comparison of TXS~0128+554 with other CSOs}

\begin{deluxetable*}{lcccccccccc}
\tablecolumns{11} 
\tabletypesize{\scriptsize} 
\tablewidth{0pt}  
\tablecaption{\label{t:othersources} Properties of {\it Fermi}-Detected and Other Nearby CSO Galaxies} 
\tablehead{\colhead{Source} & &
    & \colhead{Host}& \colhead{kpc} \\[-3ex]
    \colhead{Name} & \colhead{$z$} & \colhead{$D_L$} & \colhead{Galaxy} & \colhead{morph.}  &\colhead{LLS} & \colhead{$\theta$} &  \colhead{$\beta_{app}$}
    & \colhead{Age} & \colhead{$\nu_m$} & \colhead{f}\\[-3ex]
    \colhead{} & \colhead{} & \colhead{[Mpc]} & \colhead{} & \colhead{}
    &  \colhead{[pc]} & \colhead{[$^{\circ}$]} &\colhead{[c]} & \colhead{[y]} 
    & \colhead{[GHz]} & \colhead{}
}
\decimalcolnumbers
\startdata
\multicolumn{11}{c}{{\it Fermi}-detected CSOs} \\[2pt] \hline
TXS 0128+554 &0.036 &159  &Elliptical&C &12 &$52^{+7}_{-9}$ &$0.32\pm0.07$ &$82\pm17$ &0.66 &0.48 \\
NGC 3894 &0.011 &47  &Elliptical &E &7&10--21 &$\sim 0.1$&$59\pm5$ &5 &0.18 \\
PKS 1718$-$649 &0.014 &62  &Elliptical\tablenotemark{a} &C &2.5 &$\cdots$&$0.06 \pm 0.03$ &$70\pm30$ &3.6 &$<0.01$\\
PMN J1603$-$4904 &0.232 &1148 &Unknown\tablenotemark{b} &E  &56 &$\cdots$&$< 3$ &$>54$ &0.39 &0.37\\
\cutinhead{Low Redshift CSOs Lacking {\it Fermi} LAT Associations}
4C 31.04 &0.060 &266 & Elliptical&C & 100 &75--80&0.34 &550 &0.4 &0.016\\
PMN J1511+0518 &0.084 &378 & Elliptical& C& 11& $\cdots$&0.28 &55 &10 &0.03\\
B2 0035+22 &0.096 &435 & Elliptical& C& 22&$\cdots$ &0.5 &450 &0.4--1.4 &0.014\\
\enddata 
\tablenotetext{a}{Host has an elliptical nucleus with a prominent dust lane, surrounded by faint spiral structure.}
\tablenotetext{b}{No indications of active star formation in optical spectrum \citep{2016AA...586L...2G}.}
\tablecomments{Columns are as follows: (1) source name, (2) redshift, (3) luminosity distance in Mpc, (4) host galaxy type, (5) kpc scale radio morphology, where E = extended, and C = compact,  (6) largest projected linear size of inner jet structure as measured from hotspot-to-hotspot in pc, (7) jet viewing angle in degrees, (8) apparent expansion speed in units of the speed of light, (9) dynamical age of the radio source in years, (10) radio spectral turnover frequency in GHz, (11) ratio of core flux density to total flux density at 8 GHz. }
\end{deluxetable*}

TXS~0128+554 is the fourth confirmed CSO to be detected in $\gamma$-rays by the \fermi LAT instrument. Current CSO catalogs (e.g., \citealt{2016MNRAS.459..820T}) contain on the order of $\sim 50$ AGN, making the \textit{Fermi}-detected ones a very rare subset.   With the exception of PMN J1603$-$4904  at $z = 0.232$ (see discussion of the redshift in the Appendix), all of the \fermi CSOs are at redshift less than $z<0.04$.  In Figure~\ref{vlbi8ghzvsgamraylum} we plot 8~GHz radio luminosity versus \gr luminosity for all 4FGL AGN with a known redshift and a compact radio luminosity tabulated in the radio fundamental catalog\footnote{http://astrogeo.org/rfc/}), which is complete down to approximately 150~mJy \citep{VCS5,VCS6}.  Most GPS and CSO AGN are not detected by \fermi, yet have radio luminosities in the $10^{27}$ to $10^{28}$ W Hz$^{-1}$ range, comparable to the flat-spectrum blazars that dominate the \fermi catalogs. TXS~0128+0554 and two of the other \fermi CSOs have considerably lower radio powers, however, and lie among the lowest \gr luminosity AGN detected so far by \textit{Fermi}. 

The detection of AGN by the LAT instrument is strongly influenced by the spectral hardness (photon index $\Gamma$) of the source in the \gr regime. Weak-lined BL Lac objects tend to have harder \gr spectra (lower $\Gamma$) than flat-spectrum radio quasars, and are therefore detectable to lower flux levels. This is illustrated in Figure~\ref{plindexvsgamraylum}, where we present an updated version of the \cite{2020AA...635A.185P} plot of photon index versus \gr luminosity for all AGN in the 4LAC catalog \citep{4LAC}. The large light blue symbols represent the \fermi CSOs, with the blue cross denoting TXS~0128+554. We have also highlighted several nearby \textit{Fermi}-detected radio galaxies in the plot for comparison. Cen A is among the closest jetted AGN and along with Fornax A are the only AGN where \fermi has resolved \gr emission from the jet lobe structure \cite{2010Sci...328..725A,2016ApJ...826....1A}. M87 is a nearby well-studied jetted AGN in the Virgo cluster that lies in the same location in the $\Gamma$--$L_\gamma$ plane as Fornax A and the CSO NGC 3894.

\begin{deluxetable*}{cccccccc}
\tablecolumns{8}
\tabletypesize{\scriptsize} 
\tablewidth{0pt}
\tablecaption {\label{t:othersources2}Spectral Properties of {\it Fermi}-Detected and Other Nearby CSO galaxies}
\tablehead{\colhead{Source} & \colhead{log $L_{\nu}$} & \colhead{log $\nu L_{\nu}$} 
    & \colhead{log $\nu L_{\nu}$} & \colhead{log $\nu L_{\nu}$} & \colhead{$\Gamma $} & \colhead{log $\nu L_{\nu}$} & \colhead{$\Gamma$} \\[-3ex]
    \colhead{Name} &\colhead{(1.4 GHz)} &\colhead{(0.5--2 keV)} &\colhead{(2--10 keV)}
    &\colhead{(0.1--100\, \rm GeV)} &\colhead{(0.1--100\, \rm GeV)} &\colhead{(10--1000\, \rm GeV)} &\colhead{(10--1000\, \rm GeV)} \\[-3ex]
    \colhead{} &\colhead{[$\rm W\,Hz^{-1}$]} & \colhead{[$\rm erg\,s^{-1}$]}
    & \colhead{[$\rm erg\,s^{-1}$]} & \colhead{[$\rm erg\,s^{-1}$]} & \colhead{}
    & \colhead{[$\rm erg\,s^{-1}$]} & \colhead{}
}
\decimalcolnumbers
\startdata 
\multicolumn{8}{c}{{\it Fermi}-detected CSOs} \\[2pt] \hline
TXS 0128+554 &23.7 &42.5 &42.3 &43.2 &$2.10\pm0.09$ &42.5 &$3.1\pm0.9$\\
NGC 3894 &23.1 &40.5\tablenotemark{a} &40.8 &41.8 &$2.06\pm0.12$ &$\cdots$ &$\cdots$\\
PKS 1718$-$649 &24.3 &40.9 &41.2 &42.1 &$2.49\pm0.18$  &$\cdots$ &$\cdots$\\
PMN J1603$-$4904 &26.3 &43.5 &43.6 &45.9 &$2.02\pm0.03$ &45.6 &$2.2\pm0.10$\\
\cutinhead{Low Redshift CSOs Lacking {\it Fermi} LAT Associations}
4C 31.04 &25.3 &\textless40.6 &$\cdots$ &$\cdots$ &$\cdots$ &$\cdots$ &$\cdots$\\
PMN J1511+0518 &25.2 &$\phn$ 42.0 &42.7 &$\cdots$ &$\cdots$ &$\cdots$ &$\cdots$\\
B2 0035+22 &25.1 &$\phn\phn$41.6 &41.9 &$\cdots$ &$\cdots$ &$\cdots$ &$\cdots$\\
\enddata 
\tablenotetext{a}{0.7--2 keV luminosity.}
\tablecomments{Columns are as follows: (1) source name, (2) log of 1.4 GHz luminosity ($L_{\nu}$) in $\rm W\,Hz^{-1}$, (3) log of 0.5--2 keV x--ray luminosity  in $\rm erg\,s^{-1}$, (4) log of 2--10 keV x--ray luminosity ($\nu L_{\nu}$) in $\rm erg\,s^{-1}$, (5) log of 0.1--100 GeV $\gamma$--ray luminosity  from 4FGL in $\rm erg\,s^{-1}$, (6) 4FGL power law photon index between 0.1 and 100 GeV, (7) log of 10--1000 GeV $\gamma$--ray luminosity  from 3FHL in $\rm erg\,s^{-1}$, (8) 3FHL power law photon index between 10 and 1000 GeV. }
\end{deluxetable*}

The large luminosity of PMN J1603$-$4904 clearly sets it apart from the other CSOs and nearby radio galaxies in Fig.~\ref{plindexvsgamraylum}, as its \gr properties have more in common with the BL Lac objects. Its spectral energy distribution is strongly Compton-dominated, and its \gr emission is known to be highly variable \citep{2009ApJ...699...31A,2018AA...610L...8K}.  We have included the radio galaxy 3C 84, another highly variable \gr emitter, in the plot, as it presents an interesting case of restarted jet activity in its core, marked by large radio outbursts occurring in 1960 and 2005 \citep{2014ApJ...785...53N}. The former event led to the production of two CSO-like mini-lobes that are expanding supersonically into the surrounding medium \citep{2017ApJ...843...82K}. There has been considerable debate as to the source of the \gr rays in 3C 84, which may originate in the inner jet \citep{2018NatAs...2..472G} and/or its pc-scale lobes. Its jets show only mildly relativistic motion ($0.4 c$; \citealt{MOJAVE_XVII}), and the radio morphology, free-free absorption, and kinematics of the lobe emission suggest a high viewing angle of $60^\circ$  \citep{2017MNRAS.465L..94F}. 

Despite the relatively low levels of Doppler boosting in 3C 84 and TXS~0128+554, they both lie near the region populated by BL Lac objects in  Figure~\ref{plindexvsgamraylum}.  Along with PMN J1603$-$4904, they have substantially higher \gr luminosities than the other CSOs. It is likely that  their strong \gr emission may be associated with a heightened activity state near the base of the jet, which is associated with their recently launched jets.  A similar conclusion was reached by \cite{2020AA...635A.185P} in their comparison of NGC 3894 to other CSOs and \textit{Fermi}-detected AGN. 

In Table~\ref{t:othersources} and Table~\ref{t:othersources2} we have
compiled information on all \textit{Fermi}-detected and other
confirmed CSOs in the literature with $z < 0.1$. We provide detailed
individual descriptions of these sources in the Appendix. This is not
likely a complete list, as new CSOs (such as TXS~0128+554) are still
being discovered. In particular, we have omitted three AGN that have
been proposed as CSOs by \cite{2016MNRAS.459..820T}: VIPS J09062+4636,
VIPS J11488+5924, VIPS J12201+2916, since they do not have published
kinematic ages. We have also excluded the compact AGN OQ 208, as it
has shown a significant radio flux density decline by a factor of
$\sim$ 4.5 since 2000, which is highly atypical of CSOs (e.g.,
\citealt{2001AJ....122.1661F}). The three CSOs with $z < 0.1$ that do
not have \fermi LAT detections are 4C 31.04, PMN J1511+0518, and B2
0035+22.

Comparing the properties of the nearby non-\fermi detected CSOs, all have low pc-scale core fractions, which is typical for members of this AGN class \citep{2001AA...377..377S}. Whereas 4C 31.04 and B2 0035+22 both have kinematic ages of $\sim 500$ y, roughly an order of magnitude larger than the \textit{Fermi}-detected CSOs, PMN J1511+0518 has a relatively young age of only 55 y. It has an X-ray luminosity comparable to TXS~0128+554 and would be a strong candidate for future detection by the \fermi LAT instrument. 

\section{SUMMARY}

We have performed a {\it Chandra} X-ray and multi-frequency radio VLBA
study of the AGN TXS~0128+554, which is associated
with the \fermi \gr source 4FGL J0131.2+5547.  We summarize our
major findings as follows:

(i)  Our multi-frequency VLBA observations between 2.3 GHz and 22.2
GHz reveal a compact radio structure typical of the compact symmetric
object (CSO) class. Archival VLA and published GMRT observations
indicate no extended radio emission on kpc-scales. On pc-scales the
emission consists of a strong flat-spectrum core flanked by two
steep-spectrum lobes, with a diameter of approximately
11~pc, that extends to $\sim 16$~pc at the lowest observing
frequencies.  

(ii) The flat-spectrum core is atypically bright compared to the
lobes for a typical CSO, and shows no apparent shift in position
(upper limit = 0.03~mas) between 5.0~GHz and 22.2~GHz. The cores of
most compact blazars typically show shifts of $\sim 0.3$~mas over this
frequency range due to opacity effects. This suggests that the core
feature in TXS~0128+554 is not the optically thick
$\tau=1$ surface, but rather a recollimation or reconfinement shock.

(iii) We analyzed 7 additional 15 GHz epochs from the MOJAVE
archive, and found significant proper motions for three features over a
three year period. The bright features C5 and C8 in the approaching
(western) lobe are moving away from the core with apparent speeds of
$0.32 c \pm 0.07 c$ and $0.28 c\pm 0.06 c$, respectively. Feature C1
in the receding lobe has an apparent motion of  $0.174 c \pm 0.042 c$
toward the core. 

(iv) The relativistic expansion speeds suggest that the emission from
the approaching and receding effects suffers from time delay effects,
and are viewed at different epochs than photons from the core region
of the source. By assuming intrinsic symmetry in the lobes, we used
the apparent size and flux density asymmetry in the VLBA images to
determine a jet axis viewing angle of $52^{+7}_{-9}$ degrees, and an
intrinsic expansion speed of $0.32^{+0.08}_{-0.07} \, c$. This implies
only mild Doppler boosting factors of $\delta = 1.2$ in the
approaching jet and $\delta=0.79$ in the counter-jet.

(v) Under the assumption of a constant advance speed in the western
lobe, we derive a kinematic age of $82 \pm 17$~y, placing TXS~0128+554
among the youngest known relativistic AGN jet systems.

(vi) We detected TXS~0128+554 in our targeted 19.3~ks {\it Chandra} ACIS-S imaging
observations, and the structure is consistent with an unresolved point
source. The spectrum between 0.5 keV and 7\,keV is well fit by an
absorbed power law model with photon index $\Gamma = 2.38 \pm
0.10$. The fitted absorbing column of ${\rm
N}_H(z)=6.7\times10^{21}$\,cm$^{-2}$ is in agreement with the expected
\HI absorbing column density predicted from the $N_{\matHI}$ -- linear
size relation, as well as the trend between source size, radio
luminosity and absorbing column studied by
\cite{2019ApJ...871...71S}. We do not find evidence for a spectral
feature at 6.4 keV that would be due to the presence of a reflection
of the primary absorbed X-ray radiation off a cold medium. The
relatively soft X-ray spectrum compared to other X-ray detected CSOs
may be indicative of a thermal emission component, for which we were
able to obtain an upper temperature limit of kT = 0.08 keV.

(vii) A distinct emission gap between the bright inner jet and the
outer lobe structure, as well as the lack of compact,
inverted-spectrum hotspots indicate that TXS~0128+554 has undergone
pc-scale episodic jet activity. The jets originally became active
approximately 90 y ago, producing and energizing the outer lobe
structures. After a period of quiescence, the jets were re-launched
roughly 10 y ago, producing the bright inner jet features and
increasing the synchrotron emission from the compact core.
Intermittent jet activity has been proposed
\citep{1997ApJ...487L.135R,2009ApJ...698..840C} and observed in young
radio sources \citep{1998A&A...336L..37O}. TXS\,0128+554 is the first
CSO with a measured kinematic age which can potentially provide
constraints on the theoretical models.

(viii) The 1.4 GHz radio luminosities of TXS~0128+554 and two of three
other \textit{Fermi}-detected CSOs are in the range of $10^{23}$ W Hz$^{-1}$ to
$10^{24}$ W Hz$^{-1}$, several orders of magnitude below the typical
values for CSOs. These two \gr CSOs: NGC 3894, PKS 1718$-$649, both
have redshifts below 0.02, while the third \gr CSO, PMN J1603$-$4904
has a redshift of 0.23, based on three blended lines in a near-featureless
spectrum \citep{2016AA...586L...2G}. Comparing the properties of the
\gr CSOs to three other known CSOs with $z < 0.1$, the latter have
radio powers of $\sim 10^{25}$ W Hz$^{-1}$ and very weak/non-detected
core features. We therefore find support for the suggestion put forward by
\cite{2020AA...635A.185P} that the \gr emission in these CSOs is
originating in the core region, and is a hallmark of recently
started jet activity. Indeed, a fit to our measured properties of TXS
0128+554 based on the high-energy inverse Compton model of
\cite{2008ApJ...680..911S} predicts a \gr lobe (cocoon) luminosity
that is three orders of magnitude lower than that observed by
\fermi. 

\acknowledgments

We thank Alexander V. Plavin, Eduardo Ros, and the anonymous referee
for useful suggestions on the manuscript.  This work was supported by
NASA-\textit{Fermi} grant 80NSSC17K0517, NASA-\textit{Chandra} grant
GO9-20085X, and NASA contract NAS8-03060 (Chandra X-ray Center).  YK
and AB were supported by the Russian Science Foundation grant
16-12-10481.  The MOJAVE project was supported by NASA-\textit{Fermi}
grants NNX08AV67G, NNX12A087G, NNX15AU76G, and 80NSSC19K1579.

The Very Long Baseline Array and the National Radio Astronomy Observatory are facilities of the National Science Foundation operated under cooperative agreement by Associated Universities, Inc. This work made use of the Swinburne University of Technology software correlator \citep{2011PASP..123..275D}, developed as part of the Australian Major National Research Facilities Programme and operated under licence.
This research has made use of data from the OVRO 40-m monitoring program \cite{2011ApJS..194...29R}, which is supported in part by NASA grants NNX08AW31G, NNX11A043G, and NNX14AQ89G as well as NSF grants AST-0808050 and AST-1109911.
This research has made use of the NASA/IPAC Extragalactic Database (NED) which is operated by the Jet Propulsion Laboratory, California Institute of Technology, under contract with the National Aeronautics and Space Administration. The authors made use of the database CATS \cite{2005BSAO...58..118V} of the Special Astrophysical Observatory.
This research has made use of data obtained by the  {\it Chandra} X-ray Observatory, and software provided by the {\it Chandra} X-ray Center (CXC) in the application packages CIAO and {\it Sherpa}.

\facilities{VLBA, \textit{Chandra}, \textit{Fermi}}.

\appendix
\section{\label{fermi_csos}Low-redshift CSOs detected by \fermi}

PKS 1718$-$649: This was the first reported \gr loud CSO
\citep{2016ApJ...821L..31M}, and is located in the nucleus of the
nearby ($z = 0.014$) galaxy NGC 6328 which has a LINER type optical
spectrum \citep{1985ApJ...289..475F}. The galaxy has a bright nuclear
elliptical region, a prominent dust lane, and faint outer spiral
structure that may be indicative of an ongoing merger
\citep{2000AA...358..499F,2018AA...614A..42M}.  The AGN is embedded in
a dense nuclear gas environment, and exhibits radio spectral
variability associated with free-free absorption
\citep{2015AJ....149...74T}. It has been studied in X-rays with {\it
  Chandra} \citep{2016ApJ...823...57S} and {\it XMM-Newton}
\citep{2018AA...612L...4B}. The 8.4 GHz and 22 GHz VLBI observations
of \cite{2019AA...627A.148A} show two bright lobes with a projected
separation of only 2.5 pc. They suggest the core is located in between
these lobes, in a region with an inverted spectrum, but its precise
location could not be determined. The lack of long term flux density
variability and overall GPS spectrum led \cite{2002ApJS..141..311T} to
classify the source as a CSO, which is confirmed by
\cite{2019AA...627A.148A}.  \cite{2009AN....330..193G} reported a lobe
advance speed of 0.07 c and a kinematic age of 91 y. The multi-epoch
TANAMI observations obtain a kinematic age of ($70 \pm 30$) y, based
on the observed hot spot separation rate of $(0.13 \pm 0.06)$ mas
y$^{-1}$. Assuming symmetry, this gives a lobe advance speed of $0.06\,
c \pm 0.03\, c$.

NGC 3894: This low-luminosity CSO is located in an elliptical galaxy at $z= 0.01075$, and was found to be \gr loud by
\cite{2020AA...635A.185P} after stacking 10.8 y of Pass 8 \fermi LAT
data. The 5 GHz VLA image of \cite{1998ApJ...498..619T} shows a bright core flanked by radio lobes separated by $\sim 800$ pc. On mas scales, the flat spectrum core is flanked by two bright, continuous jets with projected lengths of $\sim 2$ pc \citep{2016MNRAS.459..820T}. Fainter diffuse emission is visible in both jets farther from the core in low-frequency VLBI images \citep{1998ApJ...498..619T}.  The inner radio source has a high core fraction ($f = 0.18$). \cite{2020AA...635A.185P} find expansion speeds of $\sim 0.1$ c and a viewing angle of $10^\circ \le \theta < 21^\circ$ for the inner jets. These measurements imply a young kinematic age of $59 \pm 5$~y.

PMN J1603$-$4904: This AGN is associated with a bright and variable
Fermi LAT \gr source with a strongly Compton-dominant spectral energy
distribution and hard photon index above 10 GeV
\citep{2018AA...610L...8K}. The source is located at low galactic
latitude, and VLT/X-Shooter spectrographic observations by
\cite{2016AA...586L...2G} revealed a blazar-like, non-stellar
optical/NIR continuum with no apparent contribution from the host
galaxy. They detected a blended H$\alpha$-[NII] complex at 8100 \AA,
and two blended [SII] lines at 8250 \AA, as well as a much fainter
emission line in the NIR. The few detectable spectral lines led them
to classify the source as a radio galaxy powered by a LINER/Seyfert
nucleus. The observed turnover in the radio spectrum at $\sim 400$ MHz
\citep{2016AA...593L..19M} and the small overall extent of the VLBI
radio emission \citep{2014AA...562A...4M} have led to a CSO
classification for the source. \cite{2016AA...593L..19M} concluded
that the source must also have larger scale low-frequency radio
emission based on comparisons with single dish flux density
measurements.  Multi-epoch VLBI observations by the TANAMI
collaboration \citep{2014AA...562A...4M} show a central feature
flanked by two hotspots with separation $\sim$ 50 pc, with continuous
jet emission in between them. They set a limit of $< 3$ c on any
possible expansion speed.

\section{\label{nonfermi_csos}Low-redshift CSOs not detected by \fermi}

4C 31.04: This AGN is located in a nearby ($z = 0.059$) giant elliptical galaxy and is interacting with dense gas associated with a 2 kpc diameter circumnuclear disk \citep{2019MNRAS.484.3393Z, 2012AA...546A..22S}. Its radio structure consists of a flat spectrum core flanked by two radio lobes, with an overall diameter of $\sim 100$ pc. \cite{2001ApJ...552..508G} estimate a viewing angle of 75 to 80 degrees for the jets from the line of sight. 
The integrated radio spectrum is peaked at $\sim 400$ MHz.
Both radio lobes have compact hotspots, and the
flat spectrum core contains a typically small fraction ($f = 0.016$) of the
total mas-scale flux density.
\cite{2003AA...399..889G} measured a lobe advance
velocity of $0.34c$ based on two VLBI epochs, and obtained a
kinematic age of 550~y.  However \cite{2019MNRAS.484.3393Z} derived a
much larger value ($\ge 12$ ky) using the bubble expansion model of
\cite{1996ApJ...467..597B}. 

PMN J1511+0518: This nearby Seyfert I galaxy ($z = 0.084$) has a simple two-sided compact radio structure, with a flat-spectrum core flanked by two hotspot-dominated lobes separated by 11 pc on the sky \citep{MOJAVE_XI}.  Low-frequency
VLBI images show additional diffuse lobe emission to the southeast of
the eastern hotspot, roughly 60 pc from the core, that may be a relic
of previous jet activity \citep{2008AA...487..885O}. The eastern hot
spot is moving steadily in a non-radial direction away from the core, at an apparent speed of $0.285 \pm 0.024$ c, according to multi-epoch 15 GHz MOJAVE VLBA observations (Lister et al., in prep.).  This yields a kinematic age of 55~y. PMN J1511+0518 has a low core fraction ($f=0.03$) and belongs to the high-frequency peaker class, with a turnover in its
integrated spectrum occurring at $\sim 10$~GHz \citep{2008AA...487..885O}. 

B2 0035+22: The radio source is hosted by a passive elliptical galaxy at $z = 0.096$ and has a compact triple radio structure 22 pc in diameter, with hotspots
in each lobe connected to the flat spectrum central core by a
continuous bridge of jet emission. The radio emission is unpolarized,
typical of the CSO class \citep{2004MNRAS.352..112B}.
\cite{2009AN....330..149P} measured an expansion velocity of $0.5c$ for
the most prominent hotspot and a kinematic age of $\sim 450$ y. The
source is not core-dominated ($f = 0.014$), and has a turnover frequency
between 0.4 and 1.4~GHz \citep{2012ApJ...760...77A}.


\bibliographystyle{aasjournal}
\bibliography{lister}

\end{document}